\documentclass[12pt]{article}
\pdfoutput=1
\usepackage{jheppub, amsmath,amssymb}

\usepackage{color,colortbl}
\usepackage[american]{babel}
\usepackage[utf8]{inputenc}
\usepackage{microtype}
\usepackage{graphicx}
\usepackage{physics}
\usepackage{bbm}
\usepackage{array}
\usepackage{tikz}
\usepackage{amscd}

\makeatletter \@addtoreset{equation}{section}
\makeatother

\newcommand{\cF}{\mathcal{F}}

\newcommand{\cK}{\mathcal{K}}

\newcommand{\cN}{\mathcal{N}}

\newcommand{\cW}{\mathcal{W}}

\newcommand{\bZ}{\mathbb{Z}}

\newcommand{\unit}{\mathbbm{1}}

\newcommand{\be}{\begin{equation}}
\newcommand{\ee}{\end{equation}}
\newcommand{\bea}{\begin{equation}\begin{aligned}}
\newcommand{\eea}{\end{aligned}\end{equation}}
\newcommand{\wt}{\widetilde}
\newcommand{\ds}{\displaystyle}

\DeclareMathOperator{\Pf}{Pf}

\newcommand{\ba}{\begin{array}}
\newcommand{\ea}{\end{array}}
\newcommand{\bpic}{\begin{tikzpicture}}
\newcommand{\epic}{\end{tikzpicture}}

\renewcommand\Im{{\mathrm{Im}}}

\newcommand{\M}{{\mathfrak M}}

\newcommand{\Llra}{\Longleftrightarrow}

\newcommand{\lra}{\leftrightarrow}

\renewcommand{\k}{\kappa}

\def\N{\nabla}

\title{Mildly Flavoring Domain Walls in $Sp(N)$ SQCD}

\author[1,2]{Sergio Benvenuti}
\author[1]{Paolo Spezzati}

\affiliation[1]{International School of Advanced Studies (SISSA), Via Bonomea 265, 34136 Trieste, Italy}
\affiliation[2]{INFN, Sezione di Trieste, Via Valerio 2, 34127 Trieste, Italy}
\emailAdd{benve79@gmail.com, pspezzat@sissa.it}

\abstract	
{We consider supersymmetric domain walls of four-dimensional $\mathcal{N}\!=\!1$   $Sp(N)$ SQCD with $F\!=\!N+1$ and $F\!=\!N+2$ flavors.

First, we study numerically the differential equations defining the walls, classifying the solutions. When $F\!=\!N+2$, in the special case of the parity-invariant walls, the naive analysis does not provide all the expected solutions. We show that an infinitesimal deformation of the  differential equations sheds some light on this issue.

Second,  we discuss the $3d$ $\mathcal{N}\!=\!1$ Chern-Simons-matter theories that should describe the effective dynamics on the walls. These proposals pass various tests, including dualities and matching of the vacua of the massive $3d$ theory with the $4d$ analysis. However, for $F\!=\!N+2$, the semiclassical analysis of the vacua is only partially successful, suggesting that yet-to-be-understood strong coupling phenomena are into play in our $3d$ $\cN\!=\!1$ gauge theories.}

\begin{document}
\maketitle

\section{Introduction}

The past decades witnessed many developments in our understanding of the infrared dynamics of strongly coupled quantum field theories (QFT) both in three and four dimensions. This progress has shown very interesting connections between the QFTs in these two different dimensions. One of the contexts in which this connection is manifest is when the four-dimensional QFTs admit domain wall solutions. These domain walls are codimension-one solitonic objects with finite tension that can be present when the vacuum structure of the model consists of multiple isolated gapped vacua.

One concrete example of this setup is offered by  Yang-Mills theory and $4d$ massive  QCD. It is believed that these theories have two gapped vacua at the special value $\theta=\pi$ of the topological theta term \cite{Gaiotto:2017yup, Gaiotto:2017tne, Creutz:2003xu, DiVecchia:2017xpu}. This setup offers the possibility of constructing a domain wall between the two vacua and studying its dynamics at low energies. Since the four-dimensional vacua are gapped, the $3d$ dynamic of the world-volume theory on the wall is decoupled from the bulk $4d$ theory. 

Notably, not only it is possible to study the IR $3d$ dynamics on the domain wall but also to connect different phases of the $3d$ theory to different low energy behavior of the $4d$ theory. For example, in the QCD case, changing the four-dimensional mass parameter of the quarks leads to a phase transition on the domain wall, from a Chern-Simon topological theory to a $\mathbb{CP}^{F-1}$ non-linear sigma model (NLSM).

Another important example with multiple gapped vacua is $\mathcal{N}=1$ massive SQCD. 
Supersymmetry allows for a special kind of domain walls: BPS domain walls preserving half of the supercharges, with computable and minimal tension. 
Since there are many vacua, and possibly more than one supersymmetric domain wall connecting each pair of vacua, the zoo of the domain walls in SQCD is considerably richer than in QCD. 

The study of the domain walls in $4d$ SQCD has been carried out in \cite{Acharya:2001dz, Chibisov:1997rc, Bashmakov:2018ghn,Dvali:1996xe,Kovner:1997ca,Witten:1997ep,Smilga:1997cx,Kogan:1997dt,Smilga:1998vs,Kaplunovsky:1998vt,Dvali:1999pk,deCarlos:1999xk,Gorsky:2000ej,Binosi:2000jb,deCarlos:2000jj,Smilga:2001yz,Ritz:2002fm,Ritz:2004mp,Armoni:2009vv,Dierigl:2014xta,Draper:2018mpj,Hsin:2018vcg,Delmastro:2020dkz}.

Acharya and Vafa \cite{Acharya:2001dz} studied the domain walls of pure SYM for $SU(N)$ gauge group, proposing an appropriate TQFT as the $3d$ effective description of the $k$-wall. SYM's with other gauge groups were considered in \cite{Bashmakov:2018ghn, Hsin:2018vcg,Delmastro:2020dkz}.

\cite{Bashmakov:2018ghn} studied domain walls in SQCD with flavors, with $SU(N)$ and $Sp(N)$ gauge group and number of flavors less than $h$, the dual Coxeter number of the gauge algebra. In this note, we add more flavors to the story of  \cite{Bashmakov:2018ghn}, focussing on the case of $Sp(N)$ gauge group with $F=N+1$ and $F=N+2$ flavors ($F$ flavors means $2F$ fundamentals). In a companion paper \cite{BS:2021b} we discuss BPS domain walls of $SU(N)$ SQCD with $N$ and $N+1$ flavors.

Our strategy to study the BPS domain walls, as in \cite{Bashmakov:2018ghn},  consists of two separate parts: a $4d$ side and a $3d$ side.

On the $4d$ side, the regime of small masses is described by an effective Wess-Zumino model, leading to BPS equations which are analyzed numerically (at large masses the Super Yang-Mills (SYM) effective description is valid, with its known domain walls  \cite{Acharya:2001dz}). We  present a classification of all solutions, both for $Sp(N)$ with $F=N+1$, and  $Sp(N)$ with $F=N+2$ flavors.\footnote{Let us mention that the BPS equations of $SU(N)$ SQCD, if the baryons are set zero, are identical to the BPS equations of $Sp(N+1)$ SQCD with the same number of flavors, so the results of this paper carry over to $SU(N)$ with $N$ and $N+1$ flavors (in \cite{BS:2021b} we also find additional solutions for $SU(N)$ SQCD, where the baryons have a non-zero profile).}

One interesting special case is the $k$-wall for $F=N+2$ and $k=\frac{N+1}{2}$. In this case, a naive analysis provides only one trivial, $Sp(F)$ invariant, solution. This is at odds with expectations from $k \neq \frac{N+1}{2}$. We address this puzzle making an infinitesimal deformation of the differential equations, which is equivalent to changing the Kähler potential, explicitly breaking the flavor symmetry. These deformed equations allow us to understand better the nature of the seemingly trivial solution found. The trivial solution is "regularized" into a combination of many different solutions. We try to analyze these solutions and their Witten indexes, without finding a complete picture. We leave the complete analysis of the classification and counting (weighted by the Witten-Index) of these class of deformed solutions to future work.

On the $3d$ side, educated guesses, similar to the ones in \cite{Bashmakov:2018ghn}, about the $3d$ effective description of the physics on the domain wall are made, in terms of $3d$ $\cN=1$  Chern-Simons-matter models with a single $Sp(k)$ gauge group and $F$ fundamental fields (see \cite{Bashmakov:2018wts, Benini:2018umh, Gaiotto:2018yjh, Benini:2018bhk, Choi:2018ohn, Benvenuti:2019ujm} for recent progress on $\cN=1$ $3d$ gauge theories). The massless theories sit at a phase transition between a set of vacua (corresponding to the domain walls at small $4d$ mass) and a single vacuum (corresponding to the domain walls at large $4d$ mass, that is $4d$ SYM). These vacua host a product of a TQFT and a NLSM. Our $3d$ proposals are argued to satisfy a non-trivial infrared duality of form $Sp(k) \lra Sp(N+1-k)$, incarnating the $4d$ equivalence between the $k$ wall and the parity-reversed $N+1-k$ wall. We stress the rationale behind such $3d$ $\cN=1$ dualities, namely their close relation with known and tested $\cN=2$ dualities.

 A check that worked well in  \cite{Bashmakov:2018ghn} is the comparison of the semiclassical vacua of the massive theory across the duality and with the $4d$ analysis. In the cases studied in this paper, such a comparison works perfectly for $Sp(N)$ with $F=N+1$, while it works only partially for $F=N+2$. More precisely,  if $F=N+2$ and $k>\frac{N+1}{2}$, the $3d$ gauge theory on the wall ($Sp(k)_{\frac{N-k+1}{2}}^{\mathcal{N}=1}$ with $N+2$ fundamentals) has additional vacua at large positive masses. Such additional vacua are not seen neither in the $4d$ analysis neither in the dual $3d$ gauge theory ($Sp(N+1-k)^{\cN=1}_{-\frac{k}{2}}$ with $N+2$ fundamentals). We ascribe such a mismatch to strong coupling effects present in the $3d$ models in such a regime, possibly similar to the ones described by \cite{Komargodski:2017keh}. The analysis of these strong coupling effects goes beyond the scope of this paper.

Contrary to the cases analyzed in this paper, where the domain wall vacua host trivial TQFT's, the domain walls of $Sp(N)$ with more than $N+2$ flavors are expected to host non-trivial TQFT's, as in \cite{Bashmakov:2018ghn}. This is because at small masses the $4d$ theory can be described by the Intriligator-Pouliot dual, which is a gauge theory, not a Wess-Zumino model. We leave the analysis of domain walls of these $4d$ SCQD's to further work.

\vspace{0.4cm}

The paper is organized as follows.

\vspace{0.1cm}

In Sec. \ref{sec: domain walls} we review some basic facts about BPS domain walls of $4d$ supersymmetric theories. 

In Sec. \ref{sec: SQCD} we study numerically the $4d$ BPS equations. Special attention is devoted to the parity-invariant walls in Sec. \ref{parity-invariant}.

 In Sec. \ref{3dTH} we propose the $3d$ effective description of the $k$-walls for $4d$ $Sp(N)$ with $F = N+1, \, N+2$ flavors, which are $3d$ $\cN=1$ $Sp(k)$ Chern-Simons-matter gauge theories.


\section{BPS domain walls of $4d$ supersymmetric theories: mini-review}
\label{sec: domain walls}

The main subject of this paper is the construction and IR characterization of domain walls in four-dimensional $\cN=1$ SQCD theories. Whenever a theory has multiple discrete vacua, one can construct extended codimension-one solitonic objects called domain walls. These configurations of the fields interpolate between the two ends of the universe in which the fields have different VEVs. We will conventionally call $x$ the coordinate orthogonal to the domain wall. The domain walls have infinite energy but finite tension. This property of the domain walls prevents them from dynamically relaxing into a unique vacuum state on the whole universe. Once the system has different vacuum configurations at the ends of the universe, the dynamics generated by the equation of motions (EOMs) or by local non-singular sources cannot evolve the system into a different configuration of the fields at $x=\pm\infty$. The class of different maps that send ($x=-\infty, x=\infty$) to the corresponding vacuum configuration of the fields is a topological property of the various sectors one can define. These sectors are identified by the VEVs of the fields at $x=\pm\infty$.

Since we will deal with four-dimensional $\mathcal{N}=1$ models, we can consider a particular type of domain walls, namely the ones that preserve half of the four real supercharges. In fact, the 4d $\mathcal{N}=1$ supersymmetry algebra admits a two-brane  charge \cite{deAzcarraga:1989mza,Dvali:1996bg}. Therefore there exist domain walls that have minimal tension within their solitonic sector.  They are called BPS domain walls \cite{Dvali:1996xe,Abraham:1990nz,Cecotti:1992rm}. These objects have some nice features due to the presence of unbroken supercharges. 
 
First of all the tension $ T$ of BPS domain walls is fixed by the ``central charge'' $Z$  that extends the $\mathcal{N}=1$ superalgebra 
\begin{equation}
T=2\abs{Z}.
\end{equation}
If the model has a WZ effective description the central charge $Z=W(v_i)-W(v_j)$ is equal to the difference of the superpotential evaluated at the two vacua $v_i$, $v_j$ at $x=\pm\infty$. So we see that the tension of a BPS domain wall does not depend on D-term; hence it is insensitive to changes of the Kähler potential. It is somehow protected and determined only by the F-terms. 

Moreover, for WZ model we have also an explicit first order differential equation to compute BPS domain wall solutions \cite{Fendley:1990zj,Abraham:1990nz}:
\begin{equation}
\label{eq:diff}
\partial_{x}\Phi^a=e^{i\gamma}\mathcal{K}^{a\bar{b}}\overline{\partial_b W},
\end{equation}
where $\Phi^a$ are the chirals of the WZ model, $\mathcal{K}^{a\bar{b}}$ is the inverse Kähler metric and $e^{i\gamma}=\frac{\Delta W}{\abs{\Delta W}}$. Note also that the trajectory of the domain wall in the W-space, that is the image of $W(\Phi^a)$ along the domain wall solution, is a straight line 
\begin{equation}
\label{eq:Wline}
\partial_{x}W=e^{i\gamma}\abs{\partial W}^2.
\end{equation}
One here should point out that the very existence of the domain walls does not depend on the D-terms \cite{Cecotti:1992rm}. In other words, it is insensitive to the choice of the Kähler metric. This will allow us, in the following, to find domain wall solutions, to choose a sensible Kähler metric, without singularities along the domain wall solution.

\section{Numerical analysis of the BPS equations}
\label{sec: SQCD}

We are interested in four-dimensional $\mathcal{N}=1$ SQCD with gauge group $Sp(N)$ and $2F$ fundamental flavors $Q$. 
The IR behavior of the models is well known \cite{Taylor:1982bp, Seiberg:1994bz, Seiberg:1994pq, Intriligator:1995ne}. If the quarks are massless, the physics at low energies crucially depends on the rank of the gauge group and on the number of flavors. Instead, if the quarks are massive, the theory always has $N+1$ distinct and massive vacua, regardless of the number of flavors. The vacua arise from spontaneous breaking of the $\bZ_{2N+2}$ R-symmetry of massive SQCD down to $\bZ_2$,  therefore they are all related by $\bZ_{N+1}$ R-symmetry rotations. Since the vacua are isolated,  BPS domain walls connecting any pair of vacua  in principle are possible. Since the vacua are related by the $\mathbb{Z}_{N+1}$ R-symmetry, the inequivalent types of BPS walls are classified by the elements of the broken part $\bZ_{N+1}$ of the R-symmetry group. In other words, there are $N+1$ different sectors of domain walls, classified by the element $k \in \bZ_{N+1}$ that relates the vacuum configuration at $x = +\infty$ to the one at $x = -\infty$.
Moreover, the domain wall sectors $k$ and $N+1-k$ are simply related by a parity transformation.

The case of $F \leq N$ has been considered in \cite{Bashmakov:2018ghn}. In this paper, we discuss in some detail the cases $F=N+1$ and $F=N+2$.

The theory has $2F$ flavors of quarks $Q_I$, where $I=1, \dots, 2F$, in the fundamental representation (the number of flavor must be even because of a global gauge anomaly \cite{Witten:1982fp}) and no superpotential. The non-anomalous continuous global symmetry is $SU(2F) \times U(1)_R$. Regarding $Sp(N)$ as the subgroup of $SU(2N)$ that leaves the $2N \times 2N$ symplectic form $\Omega = \unit_N \otimes i\sigma_2$ invariant, we indicate the flavors as $Q^\alpha_I$ with $\alpha = 1, \dots, 2N$. We  introduce the antisymmetric meson matrix $M_{IJ} = \Omega_{\alpha\beta} Q^\alpha_I Q^\beta_J$. The low energy behavior of this theory was discussed in detail in \cite{Intriligator:1995ne}.

\subsection{$Sp(N)$ with $F=N+1$}
\label{sec:domainwallsSPN+1}

For $F=N+1$, the massless theory has a moduli space of vacua. It is parametrized by a meson matrix $M_{IJ}$ which satisfies the quantum-deformed constraint%
\footnote{The Pfaffian of a $2F \times 2F$ antisymmetric matrix $M$ is $\Pf M = \frac1{2^F F!} \epsilon^{I_1 \dots I_{2F}} M_{I_1I_2} \cdots M_{I_{2F-1}I_{2F}}$ so that $\det M = (\Pf M)^2$. The variation is $\delta \Pf M = \frac12 \Pf M \cdot \Tr (M^{-1} \delta M)$. Moreover $\Pf \Omega = 1$.}
\begin{equation}
\label{eq:constraintmoduli}
\Pf M = \Lambda^{2(N+1)} \;,
\end{equation}
in terms of the dynamically-generated scale $\Lambda$. We turn on a diagonal mass term for the flavors,
\be
W_m = \frac{m_\mathrm{4d}}2 \, M_{IJ} \Omega^{IJ} \;,
\ee
where $\Omega^{IJ}$ is the symplectic form of $Sp(F)$ (in the following, we will often indicate all symplectic forms as $\Omega$, irrespective of their dimension, and will not distinguish between upper and lower indices). This explicitly breaks the $SU(2F)$ flavor symmetry to $Sp(F)$, while leaving a discrete $\bZ_{2(N+1)}$ R-symmetry unbroken, and it also lifts most of the moduli space. The mesons transform in the rank-two antisymmetric representation of $Sp(F)$. The quantum constraint (\ref{eq:constraintmoduli}) on the would-be moduli space can be implemented with a Lagrange multiplier $A$. Therefore, the low-energy physics is described by the following effective superpotential on the mesonic space:
\begin{equation}
\label{eq:effdescription sp}
W = \frac{m_\mathrm{4d}}2 M_{IJ} \Omega^{IJ} - A \, \Bigl( \Pf M-\Lambda^{2(N+1)} \Bigr) \;.
\end{equation}
The F-term equations lead to $N+1$ gapped vacua with gaugino condensation and spontaneous R-symmetry breaking $\bZ_{2(N+1)} \to \bZ_2$:
\be
\label{eq:vacuasp}
M = \wt M \, \Omega_{2F} \;,\qquad\qquad \wt M^{N+1} = \Lambda^{2(N+1)} \;,
\ee
while $A = m_\mathrm{4d} \wt M / \Lambda^{2(N+1)}$ and $\langle \lambda\lambda \rangle = \partial W / \partial \log \Lambda^{2(N+1)} = m_\mathrm{4d} \wt M$.

When the quark mass is small, $|m_\mathrm{4d}| \ll |\Lambda|$, the effective description as a Wess-Zumino model on the mesonic space is reliable. On the other hand, when the quark mass is large, $|m_\mathrm{4d}| \gg |\Lambda|$, we can integrate the quarks out first and remain with pure $Sp(N)$ SYM, with the very same $N+1$ vacua as above.

A small complication, with respect to other values of $F$, arises because the expectation value of $M$ in (\ref{eq:vacuasp}) does not depend on the mass parameter $m_\mathrm{4d}$ but only on the dynamically-generated scale $\Lambda$. If we were able to make $|M|\gg |\Lambda|$, the theory would go in a Higgsed semiclassical regime: the low energy theory would be well described by the Wess-Zumino model (\ref{eq:effdescription sp}) with the K\"ahler potential for $M$, $\cK=\Tr\sqrt{M\Omega M^{\ast}\Omega}$, induced by the canonical K\"ahler potential in terms of quarks $Q_I$. This was the situation in \cite{Bashmakov:2018ghn}. On the other hand, if we were able to make $|M| \ll |\Lambda|$, the theory would focus around a smooth point of its moduli space, and for very low energies the K\"ahler potential would essentially be the canonical one in terms of $M$ (up to rescalings) $\cK=\Tr(M\Omega M^{\ast}\Omega)$. In our case, instead, $|M| \sim |\Lambda|$ and so we do not have control over the K\"ahler potential, except for the fact that it is smooth. However it has been shown in \cite{Cecotti:1992qh} that the Cecotti--Fendley--Intriligator--Vafa index, which counts the number of BPS domain walls with signs, is independent of smooth deformations of the K\"aler potential. Therefore we assume that a smooth deformation of the K\"aler potential does not affect the existence of the domain walls we want to study. 
In summary, to find the domain solutions, we  solve the  equation \eqref{eq:diff}, making a sensible choice for the  K\"ahler metric, that is the canonical Kähler potential for the fields $M$.

\begin{figure}[t!]
\includegraphics[width=.32\textwidth]{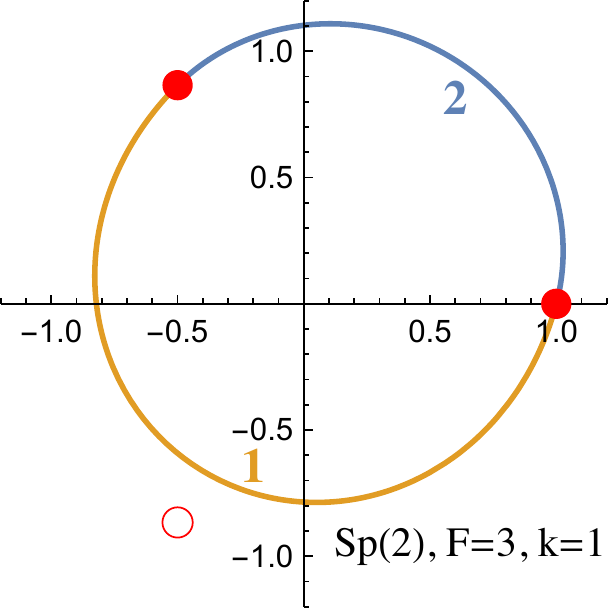} \hspace{\stretch{1}}
\includegraphics[width=.32\textwidth]{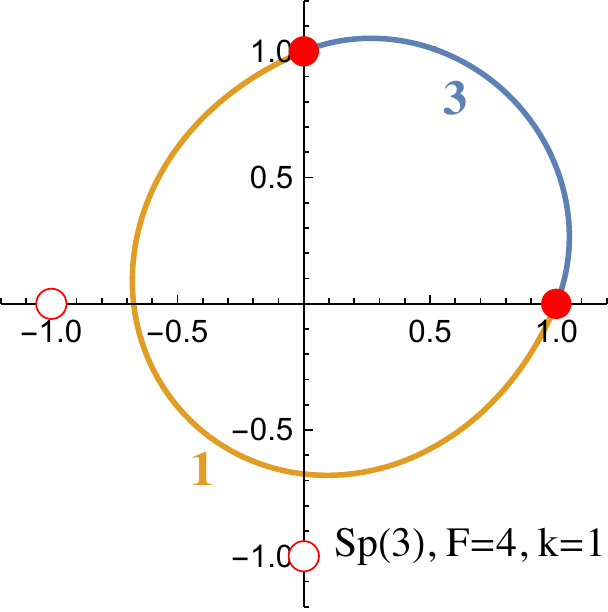} \hspace{\stretch{1}}
\includegraphics[width=.32\textwidth]{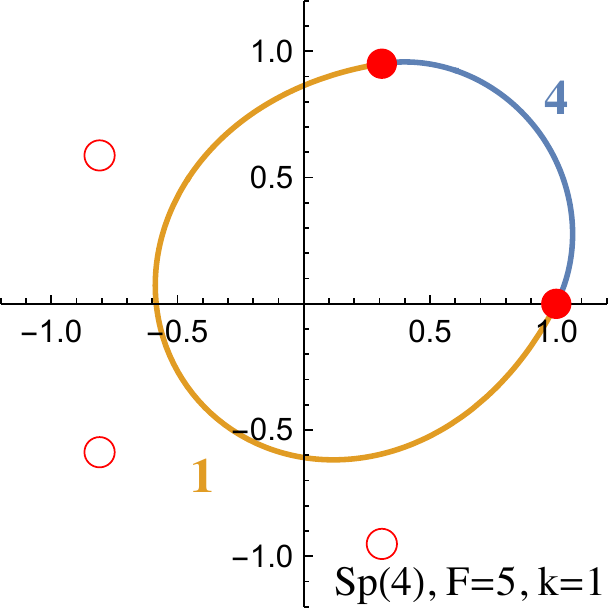}
\caption{Examples of 1-walls in $Sp(N)$ SQCD with $F=N+1$  flavors. We draw the trajectories of the eigenvalues of the meson matrix $M$ in the complex plane along the domain-wall transverse direction $x$. The filled red circles represents the expectation values of the vacua at $x=\pm \infty$, whereas the unfilled red disks represent the expectation values of the other vacua. The eigenvalues along 1-walls split into a group of $k_1 = 1$ (in yellow) and a group of $k_2 = N$ (in blue) elements. The number of eigenvalues that follow a given trajectory is indicated in the figures with a number colored like the trajectory it refers to.
\label{fig:1wallsp}}
\end{figure}

\begin{figure}[t!]
\hspace{\stretch{1}}
\includegraphics[width=.32\textwidth]{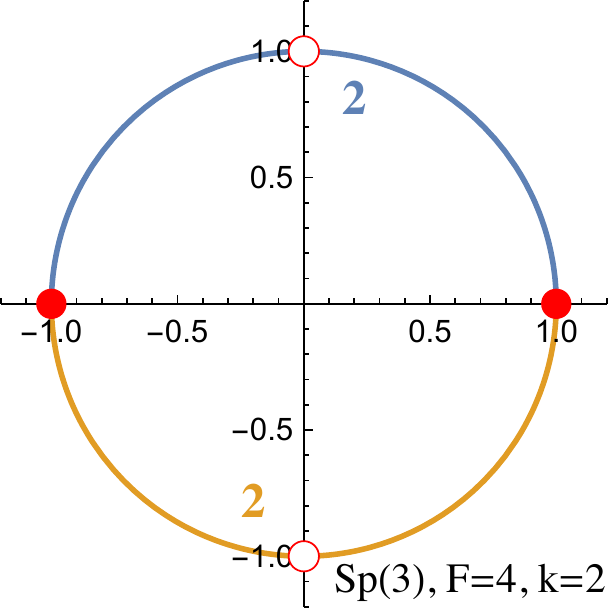} \hspace{\stretch{1}}
\includegraphics[width=.32\textwidth]{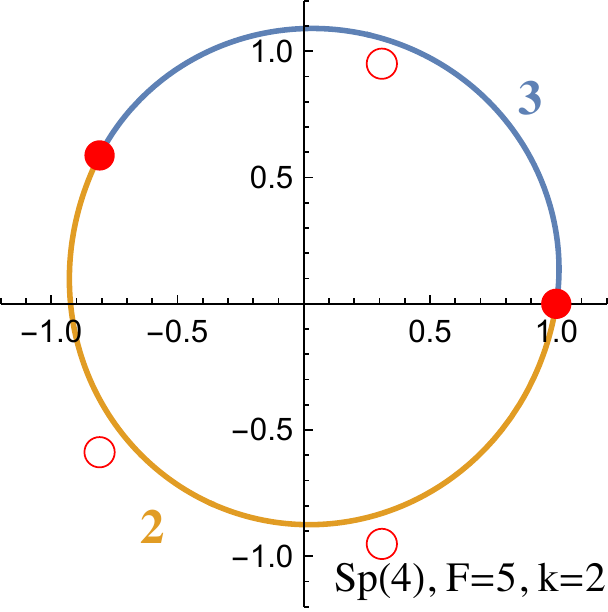} \hspace{\stretch{1}}
\caption{Examples of 2-walls in $Sp(N)$ SQCD with $F=N+1$ couples of flavors. The notation is as in Figure~\ref{fig:1wallsp}.
\label{fig:2wallsp}}
\end{figure}

Furthermore we will assume that there exist a point along the domain wall solution where the expectation value of the meson matrix is diagonalizable with the flavor symmetry, namely $M = \operatorname{diag}(\xi_1, \dots, \xi_{N+1}) \otimes i\sigma_2$. As shown in \cite{Bashmakov:2018ghn}, it follows that $M$ is diagonal everywhere on the domain wall. The problem further simplifies if we also assume that the eigenvalues split in two sets of different values $\xi_1$, $\xi_2$. In this case, we see that it is not even necessary to solve \eqref{eq:diff}, since we can find solutions by other means. Let us call%
\footnote{Here we consider the situation where both $k_{1,2}$ are non-zero. If all eigenvalues are equal, say to $\xi$, then the constraint imposes $\xi^{N+1} = \Lambda^{2(N+1)}$ leading to the $N+1$ vacua and  no domain wall solution exists.}
$k_{1,2}= 1, \dots, N$ the number of eigenvalues equal to $\xi_{1,2}$, respectively, with $k_1 + k_2 = N+1$. The superpotential takes the form
\be
W = m_\mathrm{4d} \bigl( k_1 \xi_1 + k_2 \xi_2 \bigr) - A \Bigl( \xi_1^{k_1} \xi_2^{k_2} - \Lambda^{2(N+1)} \Bigr) \;.
\ee
To simplify further, we impose the constraint, and moreover we express the meson matrix $M$ in units of $\bigl( \Lambda^{3(N+1)-F} / m_\mathrm{4d}^{N+1-F} \bigr){}^{1/(N+1)} \equiv \Lambda^2$ and set the remaining dimensionful constant $\Lambda^2 m_\mathrm{4d}$
to one. The superpotential then reduces to
\be
W = k_1 \, \xi_1 + k_2 \, \xi_1^{-k_1/k_2} \;.
\ee
As we explained in Section~\ref{sec: domain walls}, each domain wall solution traces in the complex $W$-plane a straight line connecting the values of the superpotential at the two vacua (the direction of such a line is $e^{i\gamma}$). Therefore, up to reparametrizations, the solutions can be found by simply inverting the equation
\be
W\Bigl( M \big|_{x_3 = +\infty} \Bigr) \, t + W \Bigl( M\big|_{x_3 = -\infty} \Bigr) \, (1-t) = k_1 \, \xi_1(t) + k_2 \, \xi_1(t)^{-k_1/k_2}
\ee
in terms of $\xi_1(t)$, where $t$ is some reparametrization of $x_3$.
Some examples of the solutions we found using this procedure
are sketched in Figure~\ref{fig:1wallsp} and Figure~\ref{fig:2wallsp}. It turns out that $k$-wall solutions exist for $k_1= k$ and $k_2 = N+1-k$.

The solutions we have found, in which the $N+1$ eigenvalues of $M$ split into two groups of $k$ and $N+1-k$ elements, break the flavor symmetry of the vacua according to the pattern $Sp(N+1) \to Sp(k) \times Sp(N+1-k)$. Hence, they represent a symplectic (or quaternionic) Grassmannian
\be
\mathrm{HGr}(k_1, N+1) = \frac{Sp(N+1)}{Sp(k) \times Sp(N+1-k)}
\ee
worth of domain walls. The low-energy theory on the domain walls is given by a 3d $\cN=1$ NLSM of Goldstone fields with the Grassmannian as target. We summarize the $k$-wall solutions we have found in Table \ref{table DW theories Sp F=N+1}.
\begin{table}
\begin{center}
\begin{tabular}{|c|c|c|}
\hline
Wall& Effective theory & Witten index\\
\hline
$k $& $\mathrm{HGr}(k, N+1) $& $\ds \binom{N+1}{k} \rule[-1.4em]{0pt}{3.3em} $\\
\hline
\end{tabular}
\end{center}
\caption{\label{table DW theories Sp F=N+1}Domain wall solutions found for 4d $\cN=1$ $Sp(N)$ SQCD with $F=N+1$ flavors in the regime when $m_{4d}\ll\Lambda$. For each $k$-wall sector are included also the various contributions to the Witten Index of the low energy theory on the domain wall from each solution.}
\end{table}

The solutions we found rely on the assumption that the eigenvalues split into at most two groups. We were not able to find solutions with splitting into more than two groups\footnote{Assuming that the Kähler potential is the canonical one for the fields $M_{IJ}$, the equations \eqref{eq:diff}  we have to study for the eigenvalues of the meson matrix are
\be
\partial_x\xi_i=e^{i\gamma}\biggl[\biggl(1-\frac{\prod_{k\neq i}\abs{\xi_k}^2}{D}\biggr)\biggl(\frac{\xi_i^{\ast}\prod_{j}\xi^{\ast}_j-1}{\xi_i^{\ast}\prod_{j}\xi^{\ast}_j}\biggr)+\sum_{j\neq i}\biggl(\frac{\xi_i\xi_j^{\ast}\prod_{h\neq i, j}\abs{\xi_h}^2}{D}\biggr)\biggl(\frac{\xi_j^\ast\prod_{k}\xi^{\ast}_k-1}{\xi_j^\ast\prod_{k}\xi^{\ast}_k}\biggr)\biggr],
\ee
where $D=\prod_k\abs{\xi_k}^4+\frac12\sum_{k\neq j}\abs{\xi_k}^2\abs{\xi_j}^2$. These equations have been obtained first evaluating the constraint $\prod_k\xi_k=1$, expressing the $\xi_{N+1}=\frac{1}{\prod_{i=1}^{N}\xi_i}$. Then substituting the expression for $\xi_{N+1}$ into the   superpotential \eqref{eq:effdescription sp} appropriately rescaled --- obtaining the expression $W=\sum_{i=1}^{N}\xi_i+\prod_{j=1}^N\frac{1}{\xi_j}$ --- and into the Kähler potential $\cK=\sum_{i=1}^N\abs{\xi_i}^2+\prod_{i=1}^N\frac1{\abs{\xi_i}^2}$.}. However, finding such solutions requires solving ODEs, which is a much more difficult task and we might have missed solutions. 

A  check of the completeness of our set of solutions comes from Witten indices of the low energy theories living on the domain walls at large mass, which are \cite{Bashmakov:2018ghn} the TQFT's 
\be
3d \quad Sp(k)_{N-\frac{k-3}{2}}^{\cN=1},
\ee
and have Witten Index  $\smqty(N+1\\k)$.  The  Witten Index of the TQFT (valid at large masses) is equal to the  Witten Index of the NLSM we found here (valid at small masses). See Table \ref{table DW theories Sp F=N+1}. (See also \cite{Bashmakov:2018ghn,Delmastro:2020dkz} for computation of Witten Indexes in TQFT's and in NLSM's on Grassmannians.) 

In Sec. \ref{SPN+1} we discuss a $3d$ $\cN=1$ SCFT describing the phase transition between the TQFT vacuum and the NLSM vacuum.

\subsection{$Sp(N)$ with $F=N+2$}\label{sec:SPNFN+2}

Let us now move to $F= N+2$.\footnote{We consider only $k \neq \frac{N+1}{2}$ at first. The parity-invariant case $k = \frac{N+1}{2}$ requires a special procedure, discussed in Sec. \ref{parity-invariant}.}  In this case, the low-energy $4d$ physics has a weakly-coupled  description \cite{Seiberg:1994bz, Intriligator:1995ne} in terms of a Wess-Zumino model of chiral multiplets $M_{IJ} = - M_{JI}$, with $I,J=1, \dots, 2F$ and superpotential
\begin{equation}
\label{SuperSpN+2}
W=-\frac{1}{\Lambda^{2N+1}} \Pf M \;.
\end{equation}
In the UV description  $M_{IJ}$ is the meson matrix $\Omega_{\alpha\beta} Q_I^\alpha Q_J^\beta$. The moduli space of the Wess-Zumino model is parametrized by antisymmetric matrices $M_{IJ}$ with $\rank M \leq N$, which coincides with the classical constraint in the UV SQCD theory. Adding a diagonal mass term, the IR superpotential becomes
\begin{equation}
\label{eq:Wseiberg}
W = \frac{m_\mathrm{4d}}2 \, M_{IJ} \Omega^{IJ} - \frac 1{\Lambda^{2N+1}} \Pf M \;.
\end{equation}
In the massive theory, the moduli space reduces to $N+1$ gapped vacua 
\begin{equation}
M = \wt M \, \Omega_{2F} \;,\qquad\qquad \wt M^{N+1} = m_\mathrm{4d} \Lambda^{2N+1} \;.
\end{equation}
Notice that in this case $M \to 0$ as $m_\mathrm{4d} \to 0$, so in the small mass limit the IR physics is well described by the Wess-Zumino model (\ref{eq:Wseiberg}) with canonical K\"ahler potential in terms of $M$, $\cK=\Tr(M\Omega M^{\ast}\Omega)$.


To find the domain wall solutions, we study the differential equations \eqref{eq:diff}, with superpotential given by \eqref{eq:Wseiberg} and  Kähler potential $\cK=\Tr(M\Omega M^{\ast}\Omega)$. In order to simplify the equations, we express $M$ in units of $\bigl( m_\mathrm{4d}^{F-N-1} \Lambda^{3(N+1)-F} \bigr){}^\frac1{N+1}$ and we set $\bigl( m_\mathrm{4d}^F \Lambda^{3(N+1)-F} \bigr){}^\frac1{N+1}=1$.

We make a diagonal ansatz for the meson matrix: 
\be M = \operatorname{diag}(\xi_1, \dots, \xi_F) \otimes i\sigma_2\,.\ee

With this ansatz, the "off-diagonal" differential equations  are automatically satisfied, as in \cite{Bashmakov:2018ghn}, and we are left with the "diagonal" equations.  

In order to write the $F$ complex equations for the $F$ complex eigenvalues $\xi_i$, we pass to polar coordinates. Expressing the eigenvalues in polar form, $\xi_j = \rho_j \, e^{i\phi_j}$  the $2F$ real differential equations read

\bea
\label{eq:diffpolar}
&\partial_x \rho_i=-\bigl(\prod_{j\neq i}\rho_j\bigr)\cos(\sum_{j=1}^{N+2}\phi_j-\gamma)+\cos(\phi_i-\gamma),\\
&\partial_x \phi_i=\rho_i^{-1}\bigl(\prod_{j\neq i}\rho_j\bigr)\sin(\sum_{j=1}^{N+2}\phi_j-\gamma)-\rho_i^{-1}\sin(\phi_i-\gamma)\\
\eea

These differential equations can be seen as the Hamiltonian system 
\be \rho_i=\frac1{\rho_i}\pdv{H}{\phi_i},\qquad \phi_i=-\frac1{\rho_i}\pdv{H}{\rho_i}\,,\ee 
whose Hamiltonian is\footnote{The Hamiltonian can also be written as $H=\Im(e^{-i\gamma}W(\xi_i))$. Here the Poisson tensor is not the canonical one, but it is  $J=\text{diag}(\rho_1^{-1}\otimes i\sigma_2,\dots,\rho_{N+2}^{-1} \otimes i\sigma_2)$ and the  reduced superpotential $W(\xi_i)$
\be W(\xi_i) =\sum_{i=1}^{N+2} \xi_i  - \prod_{i=1}^{N+2}\xi_i \;. \ee} 
\begin{equation}
H=-\bigl(\prod_i\rho_i \bigr)\sin(\sum_i \phi_i-\gamma)+\sum\rho_i \sin(\phi_i-\gamma).
\end{equation}

The solutions of the differential equations \eqref{eq:diffpolar}, that we found numerically, split the eigenvalues into at  most two sets: $J$ plus $F-J$.

\begin{figure}[]
\centering
\hspace{\stretch{1}}
 \includegraphics[width=.32\textwidth]{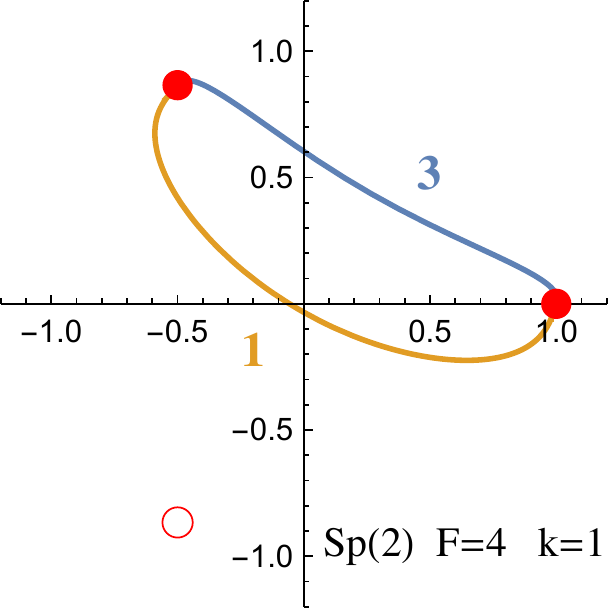} \hspace{\stretch{1}}
\includegraphics[width=.32\textwidth]{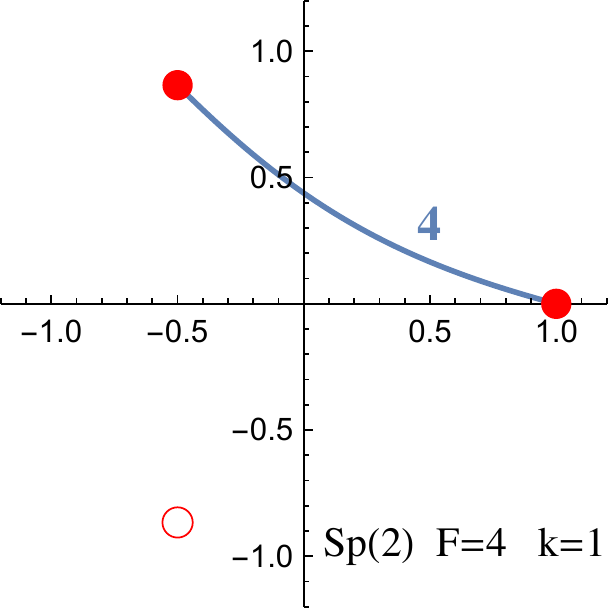} \hspace{\stretch{1}}
 \caption{Examples of 1-wall in $Sp(2)$, $F=4$.}
 \label{fig:1wallsp21}
\end{figure}

\begin{figure}[]
\centering
\hspace{0.25cm}
 \includegraphics[width=.32\textwidth]{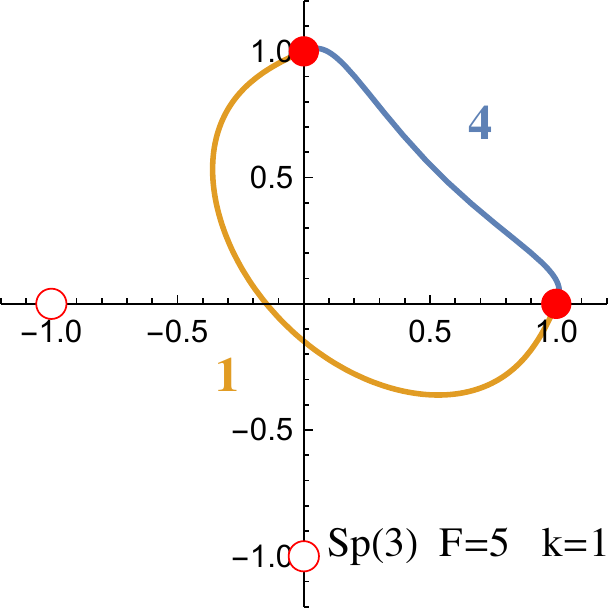}\hspace{1.5cm}
  \includegraphics[width=.32\textwidth]{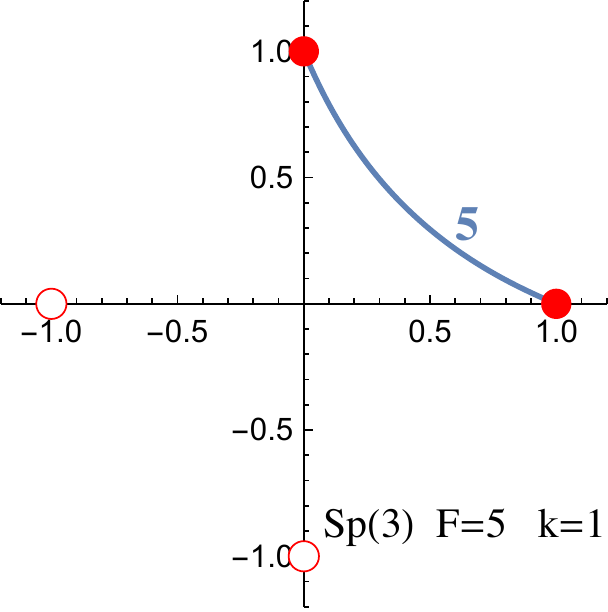}\hspace{4cm}\\
 \caption{Examples of 1-wall solutions of $Sp(3)$, $F=5$.}
 \label{fig:1wallsp31}
\end{figure}

\begin{figure}[]
\centering
\hspace{0.25cm}
 \includegraphics[width=.32\textwidth]{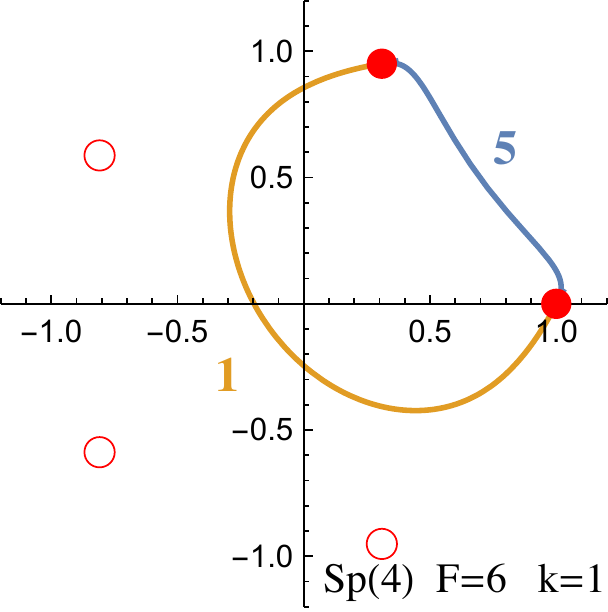}\hspace{1.5cm}
 \includegraphics[width=.32\textwidth]{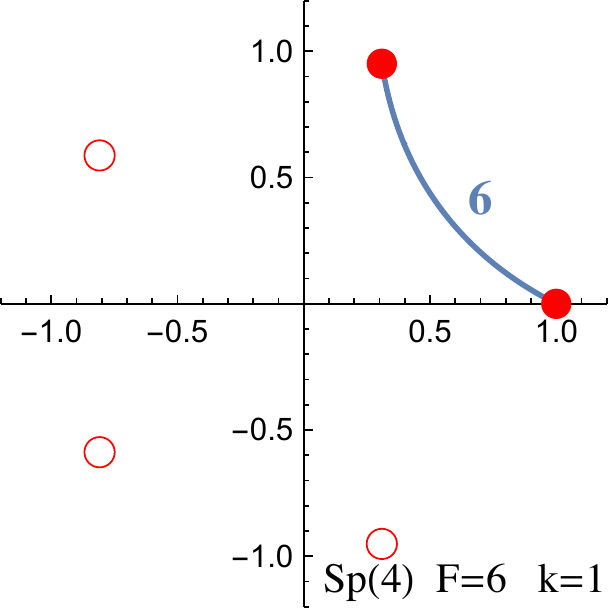}\hspace{5cm}\\
   \includegraphics[width=.32\textwidth]{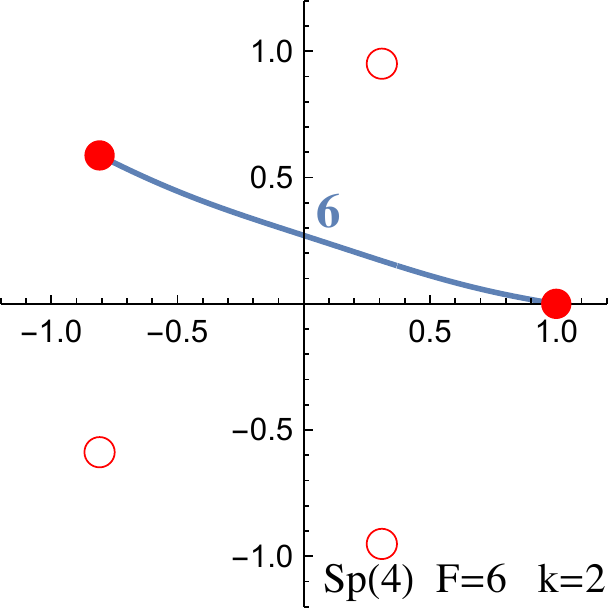}\hspace{\stretch{1}}
    \includegraphics[width=.32\textwidth]{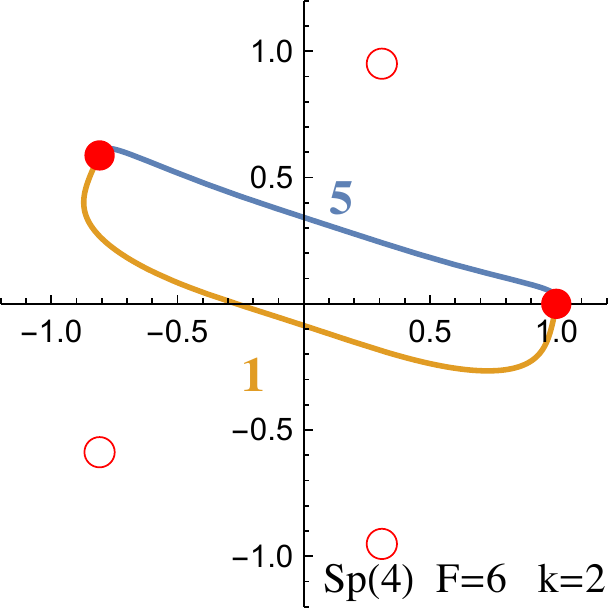}\hspace{\stretch{1}}
  \includegraphics[width=.32\textwidth]{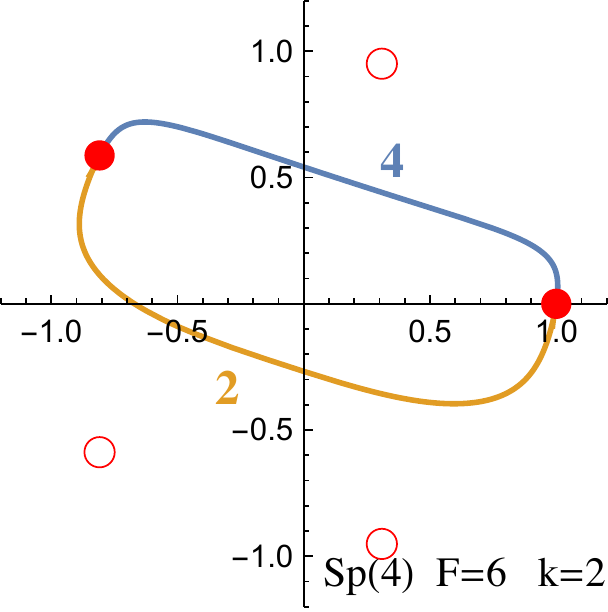}\hspace{\stretch{1}}
 \caption{Examples of domain walls in $Sp(4)$, $F=6$: on the first row there are the 1-wall solutions, whereas on the second row there are the 2-wall solutions.}
 \label{fig:1wallsp41}
\end{figure}

We plot the solution for $N \leq 4$ in Figure~\ref{fig:1wallsp21}, Figure~\ref{fig:1wallsp31} and Figure~\ref{fig:1wallsp41}. We only display $k<\frac{N+1}2$. The domain wall sector $k=\frac{N+1}2$ will be treated in sec. \ref{parity-invariant}. The cases $k>\frac{N+1}2$ are the parity reversed of $N+1-k<\frac{N+1}2$.

\

We find a $k$-wall solution for any $J=0,\dots,k$, so there are $k+1$ different solutions.
 These solutions break the $Sp(F)$ flavor symmetry to $Sp(J)\times Sp(F-J)$. Therefore, these are families of solutions parametrized by the symplectic Grassmannian 
 \be \mathrm{HGr}(J,F=N+2)\,.\ee
  The low energy theory on the domain walls is given by a 3d $\cN=1$ NLSM of Goldstone fields with target $\mathrm{HGr}(J, N+2)$. 
We sum up the various k-walls we have found in Table \ref{table DW theories Sp F=N+2}.

 \begin{table}
\begin{center}
\begin{tabular}{|c|c|c|}
\hline
Wall&Effective theory& Witten Index\\
\hline
$k$& $\mathrm{HGr}(J,N+2),\quad J\in \{0,\dots,k\}$&$   \mqty(N+1\\k)=\sum_{j=0}^{k}(-1)^{j+k}\mqty(N+2\\j)$\\
\hline
\end{tabular}
\end{center}
\caption{\label{table DW theories Sp F=N+2}The $k+1$ domain wall solutions of $4d$ $\cN=1$ $Sp(N)$ SQCD with $F=N+2$ flavors in the regime $m_{4d}\ll\Lambda$. On the right we show how the various contributions to the Witten index from each solution sum up the Witten index of the pure $Sp(N)$ SYM.}
\end{table}


The solutions found have the property that the eigenvalues of the meson matrix split into two groups, and not more. We were not able to find solutions where the eigenvalues split into three or more groups. 

A check that the solutions we found are the full set of solutions comes from the Witten Index. The alternating sum\footnote{See \cite{Bashmakov:2018wts} for the explanation for the alternating sign of the sum. This is due to the number of fermions with negative mass that are integrated out.}  of Witten indices in the $k$ sector of  Table \ref{table DW theories Sp F=N+2} coincides with the Witten Index of the k-wall of pure $Sp(N)$ Super-Yang-Mills.

\begin{figure}[]
\centering
\hspace{\stretch{1}}
 \includegraphics[width=.32\textwidth]{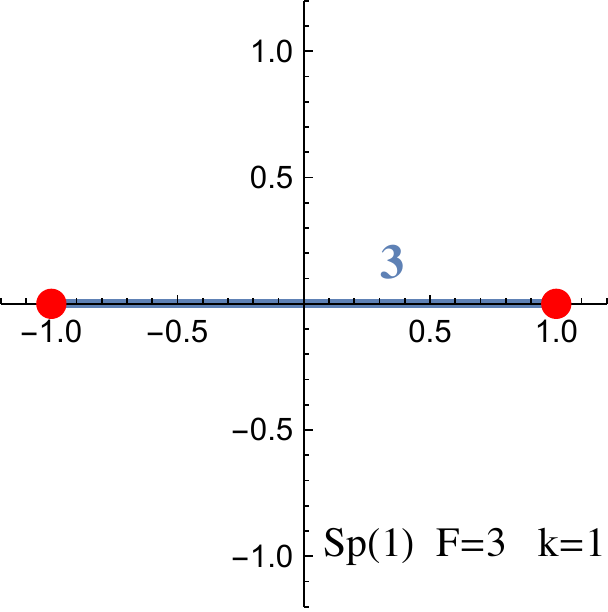}\hspace{\stretch{1}}
 \includegraphics[width=.32\textwidth]{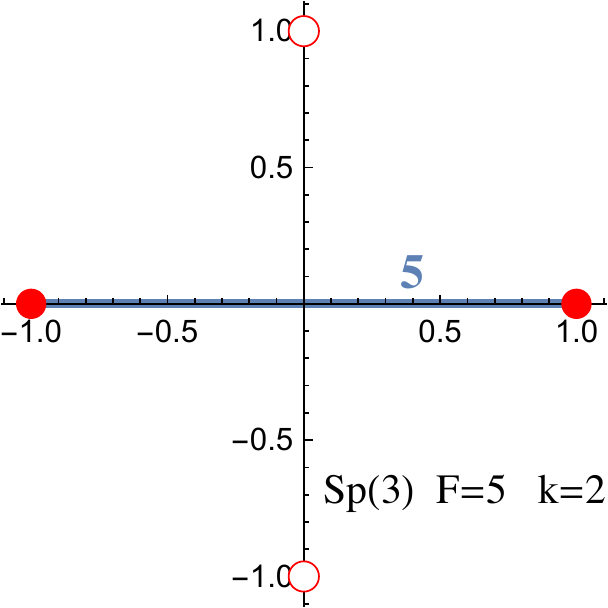}\hspace{\stretch{1}}
 \caption{Examples of $k=\frac{N+1}2$ domain walls : on the left a 1-wall solution of $Sp(1)$ with $F=3$ flavors, on the right a 2-wall solution of $Sp(3)$ with $F=5$ flavors.}
 \label{fig:1wallspline}
\end{figure}

\subsubsection{The parity-invariant walls, $k=\frac{N+1}{2}$}\label{parity-invariant}
If $N$ is odd, the $k=\frac{N+1}{2}$ wall exists and must be equivalent to its $4d$ parity transformed. For this reason we dub such domain walls \emph{parity-invariant}.

In this case, a naive numerical analysis yields only a single domain wall, the one with all the $N+2$ eigenvalues following the same trajectory (so  the global $Sp(F)$ symmetry is unbroken), which is an horizontal  straight line connecting the vacuum $M=+  \Omega_{2F}$ to the vacuum $M=-  \Omega_{2F}$ along the real line (see Figure \ref{fig:1wallspline}). This fact is in contrast with expectations from the other $k$-walls, with $k < \frac{N+1}{2}$, where we find $k+1$ different solutions (parameterized by $J=0,1,\ldots,k$ splitting the $F$ eigenvalues into $J$ and $F-J$). Analogously, for $k > \frac{N+1}{2}$, the $k$-wall, being the parity transformed $N+1-k<\frac{N+1}{2}$ wall, admits $N+2-k$ solutions. So it is natural to expect $\frac{N+1}{2}+1$ solutions for the parity-invariant walls of $Sp(N)$ with $N+2$ flavors, not just a single solution. \footnote{Other more exotic options are of course possible.}

\begin{figure}[]
\centering
\hspace{\stretch{1}}
 \includegraphics[width=.32\textwidth]{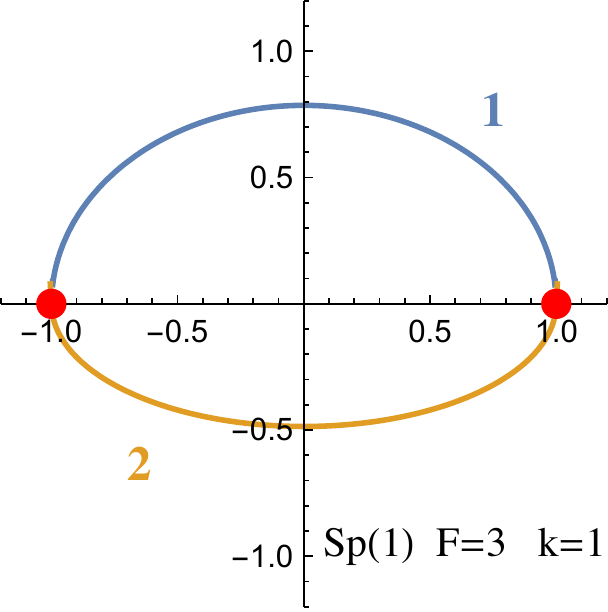}\hspace{\stretch{1}}
 \includegraphics[width=.32\textwidth]{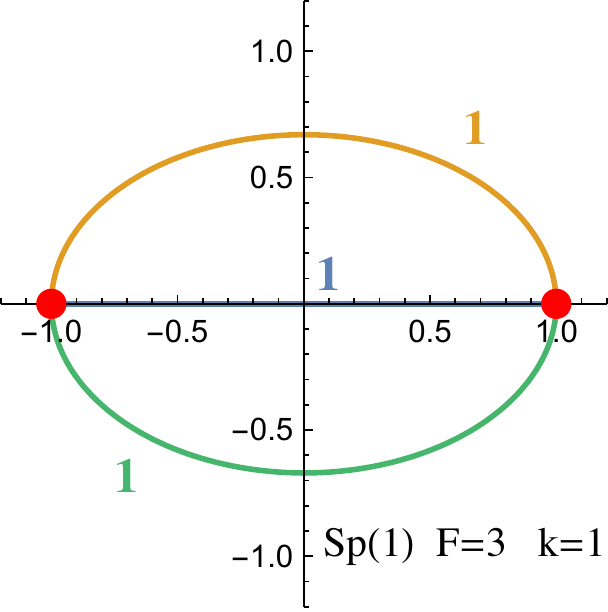}\hspace{\stretch{1}}
 \caption{Examples of deformed domain walls solutions in  $Sp(1)$, $F=3$: on the left the parameters of the deformation \eqref{eq:deformation} are $\epsilon_i=(0.005,0.005,0)$. On the right the parameters are $\epsilon_i=(0,-1,-1)$. The parameters of the right figure cannot be considered small, but they have been chosen such that the figure could be easy to read. There is also a third solution, corresponding to the charge-conjugated of left-figure, which we do not display.}
 \label{fig:1wallsp1}
\end{figure}

\begin{figure}[]
\centering
\hspace{\stretch{1}}
 \includegraphics[width=.32\textwidth]{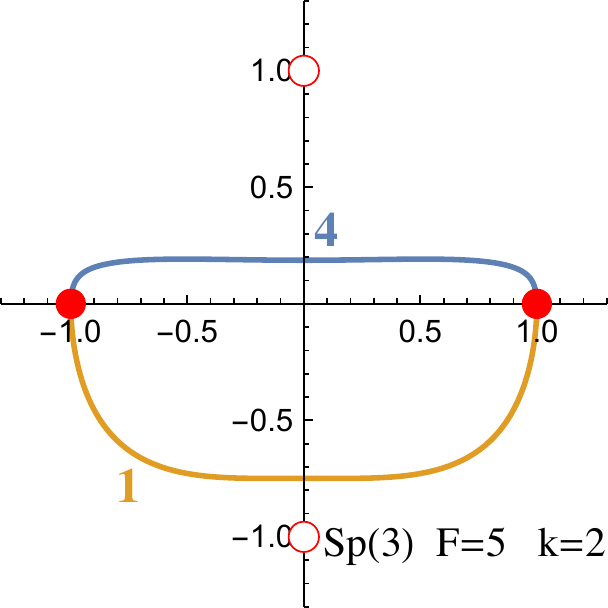}\hspace{\stretch{1}}
 \includegraphics[width=.32\textwidth]{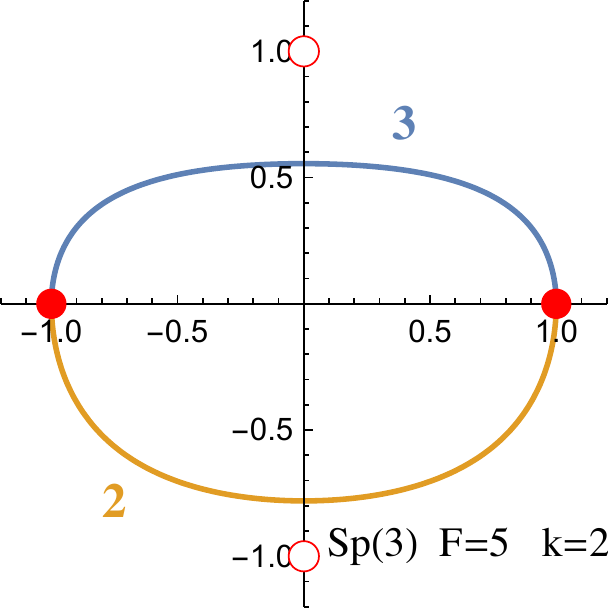}\hspace{\stretch{1}}\\
  \includegraphics[width=.32\textwidth]{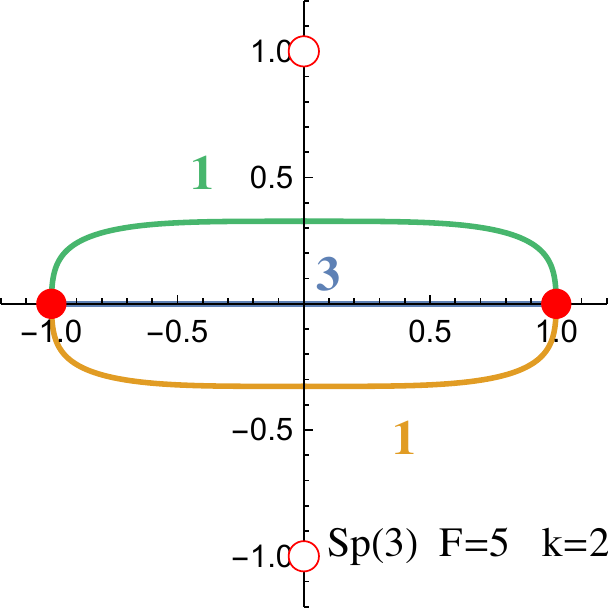}\hspace{\stretch{1}}
   \includegraphics[width=.32\textwidth]{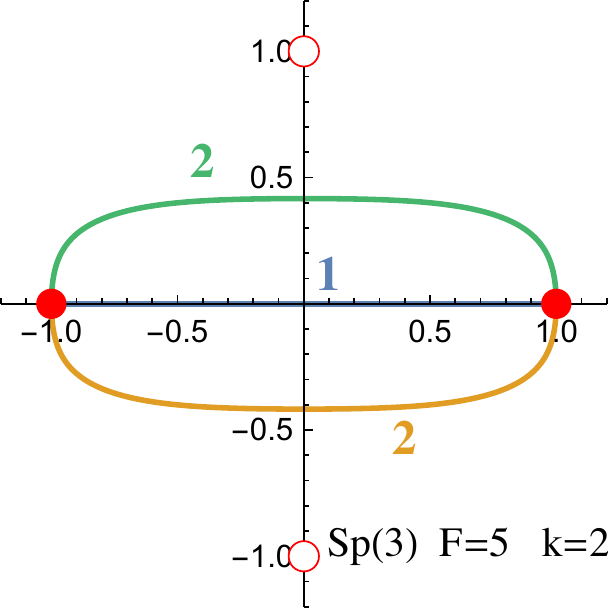}\hspace{\stretch{1}}
    \includegraphics[width=.32\textwidth]{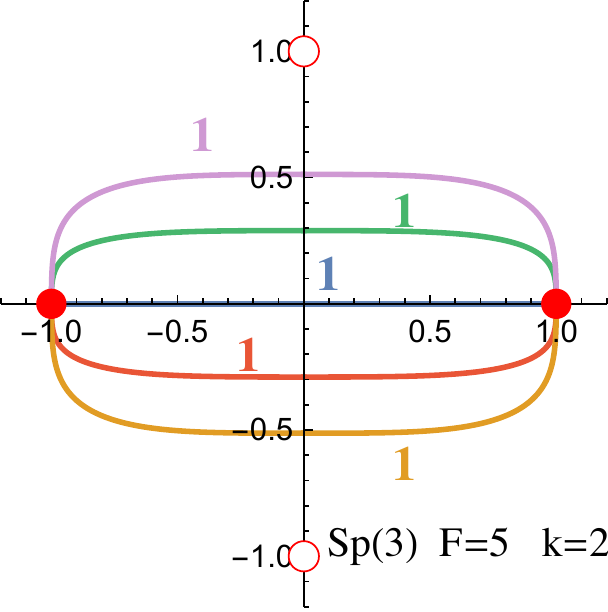}\hspace{\stretch{1}}
 \caption{Examples of domain walls in $Sp(3)$, $F=5$: the parameters of the deformations are, from left to right, top to bottom, $(0,-1,-1,-1,-1,-1)$, $(-1,-1,-2,-2,-2)$, $(0.01,0.01,-0.1,-0.1,-0.1)$, $(0.01,0.01, 0.01,0.01,-0.1)$, $(0.1,0.1,0,0.2,0.2)$.}
 \label{fig:2wallsp3}
\end{figure}

\begin{figure}[]
\centering
\hspace{\stretch{1}}
 \includegraphics[width=.32\textwidth]{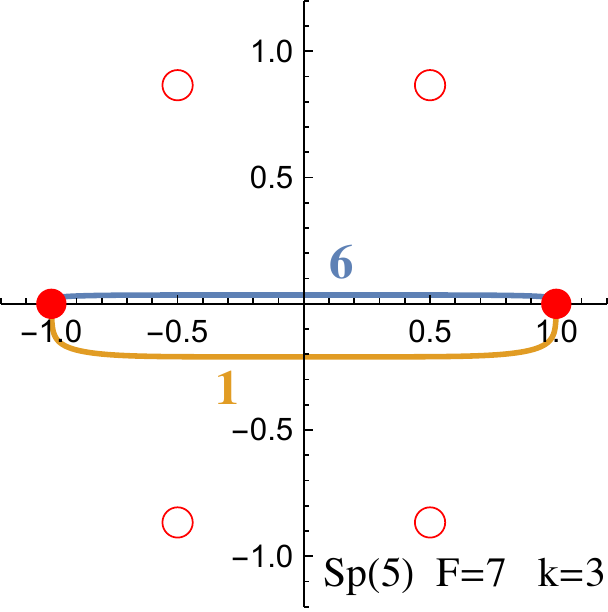}\hspace{\stretch{1}}
 \includegraphics[width=.32\textwidth]{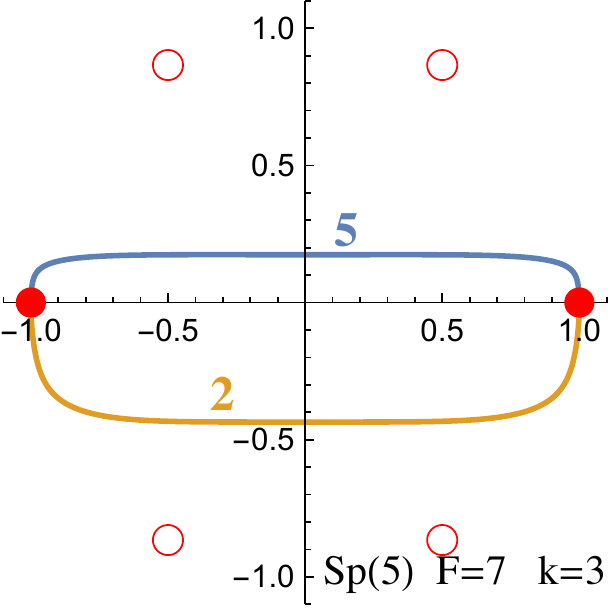}\hspace{\stretch{1}}
  \includegraphics[width=.32\textwidth]{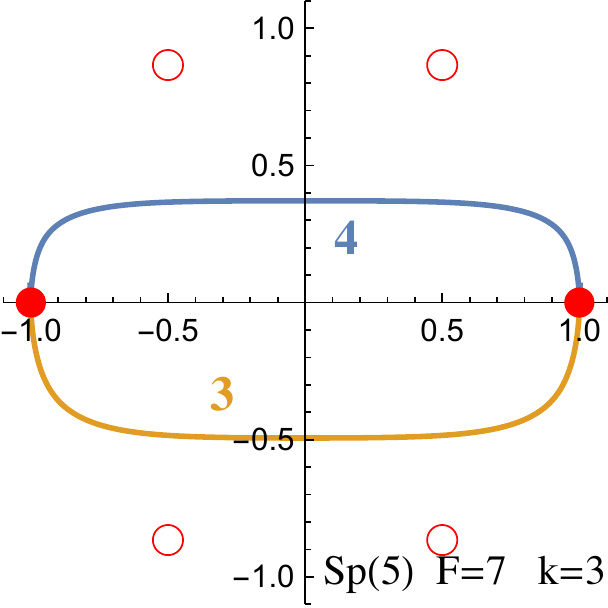}\hspace{\stretch{1}}\\
  \includegraphics[width=.32\textwidth]{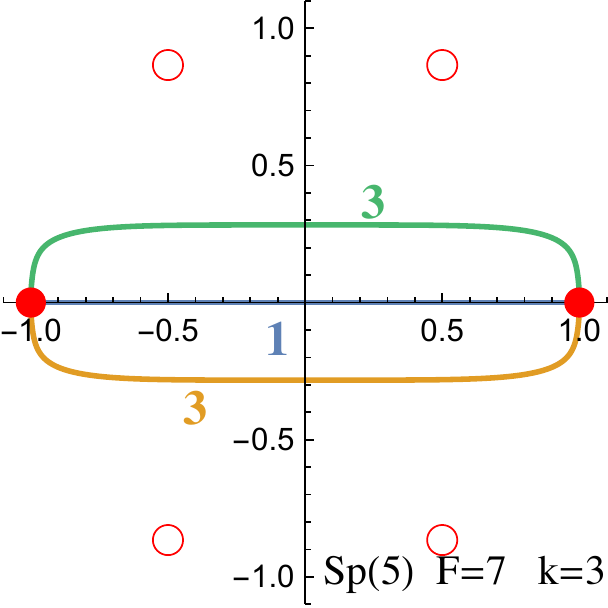}\hspace{\stretch{1}}
    \includegraphics[width=.32\textwidth]{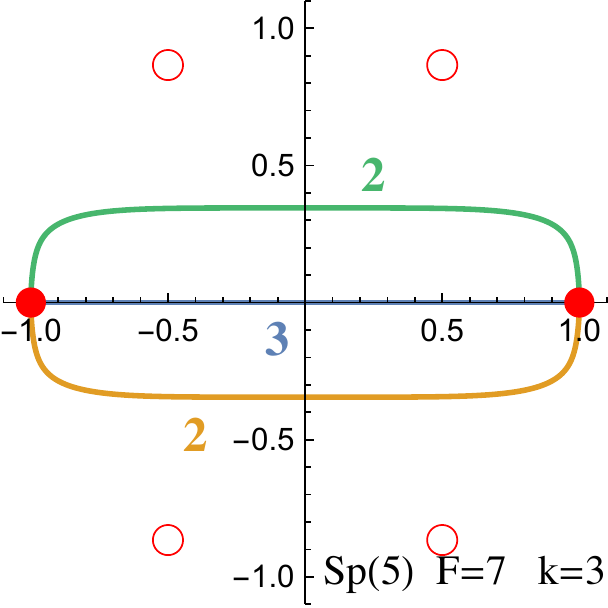}\hspace{\stretch{1}}
      \includegraphics[width=.32\textwidth]{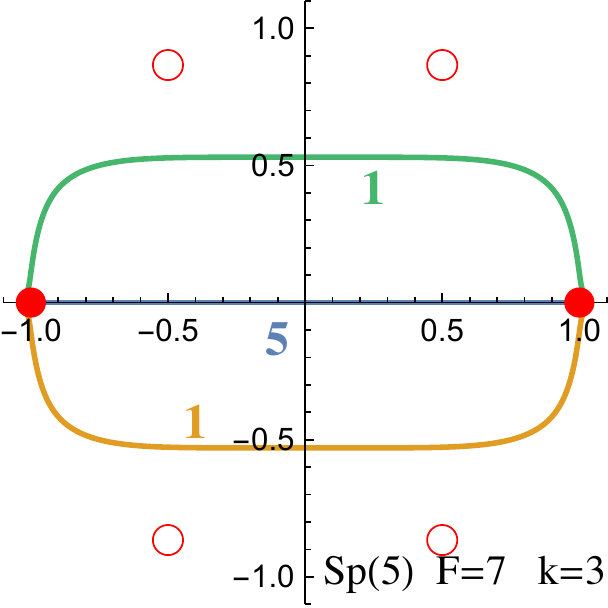}\hspace{\stretch{1}}\\
    \includegraphics[width=.32\textwidth]{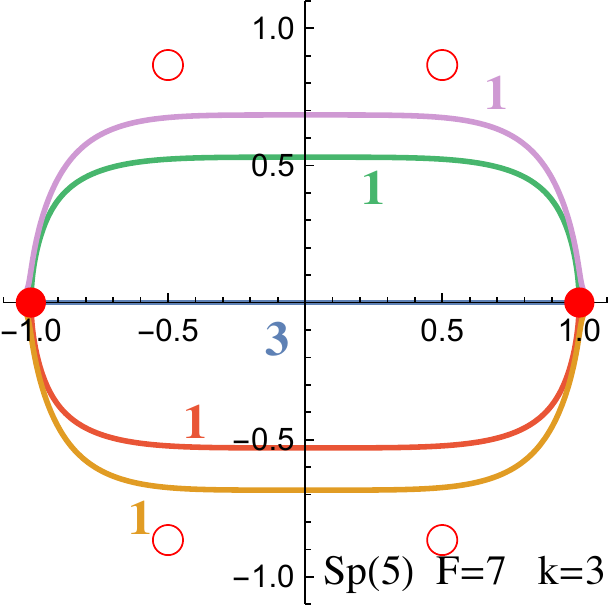}\hspace{\stretch{1}}
       \includegraphics[width=.32\textwidth]{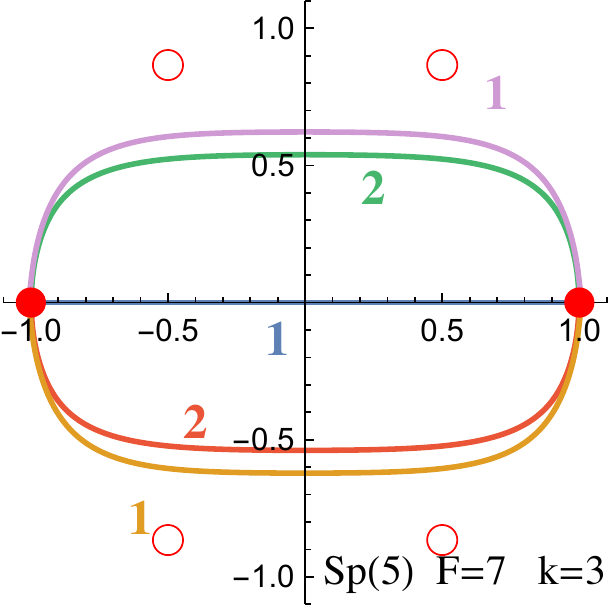}\hspace{\stretch{1}}
          \includegraphics[width=.32\textwidth]{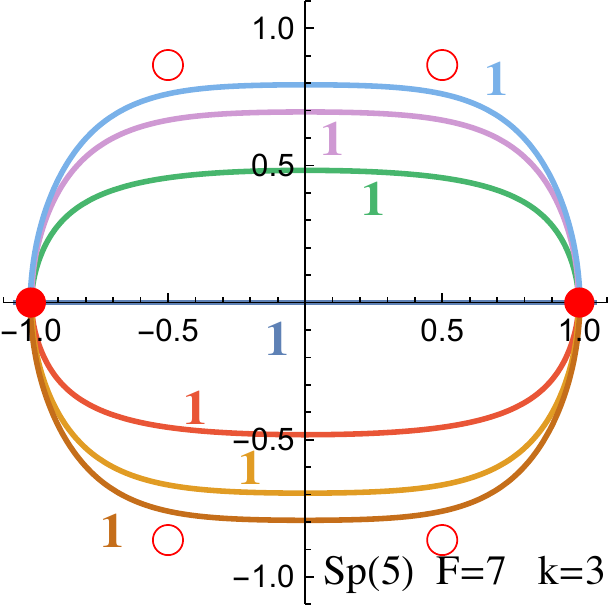}\hspace{\stretch{1}}
 \caption{Examples of domain walls in $Sp(5)$, $F=7$: here are listed the parameters of the deformations used to compute these solutions. First line from left to right, the $\epsilon_i$ are $(0,-0.1,-0.1,-0.1,-0.1,-0.1,-0.1)$, $(0.1,0.1,-0.1,-0.1,-0.1,-0.1,-0.1)$, $(0.1,0.1,0.1, -0.1,-0.1 -0.1,-0.1)$. Second line from left to right, $(0,0.1,0.1,0.1,0.1,0.1,0.1)$, $(0,0,0,0.1,0.1,0.1,0.1)$, $(-0.1,-0.1,-0.1,-0.1,-0.1,0.1,0.1)$. Third line from left to right $(0.1,0.1,0.2,0.2,-0.1,-0.1,-0.1)$, $(0.05,0.05,0.05,0.05,-0.1,0.1,0.1)$, $(0.05,0.05,0.2,0.2,0.3,0.3,-0.1)$.}
 \label{fig:3wallsp5}
\end{figure}

In this subsection we give our interpretation of this puzzle.

In the case of the parity-invariant wall, we found that upon making a small deformation of the system of ODE's, more solutions appear. All these additional solutions collapse to the straight line solution if we tune the deformation to zero. Notice that this is not true for other $k$'s: generically, a small deformations does not generate additional solutions on top of the ones discussed previously. One can think of such small deformation as a regularization of the problem of finding and counting the solutions of the system of differential equations.

The deformations we are considering are equivalent to a deformation of the K\"ahler potential that break the global $Sp(F)$ symmetry to a product of smaller $Sp$ factors. We break the global symmetry explicitly in order to resolve the degeneracy of the "real solutions".\footnote{One might try to consider other deformations of the K\"ahler potential that do not break the global $Sp(F)$ symmetry, e.g. higher order terms. We expect such deformations to change the shape of the solutions, but not to change the number of solutions of the undeformed differential equations.} The deformations can be parametrized by the block-diagonal matrix $J=\text{diag}(\epsilon_1,\dots,\epsilon_{N+2})\otimes i\sigma_2$. The variation of the K\"ahler potential is
\be
\label{eq:deformation}
\delta \cK=\frac14 \Tr(MJ)\Tr(M^{\ast}J).
\ee
Note that when $\epsilon_i\neq \epsilon_j$ for some $i$ and $j$ then the $Sp(F)$ is explicitly broken. 

 The solutions we found are depicted in Figure \ref{fig:1wallsp1} ($1$-wall of $Sp(1)$), Figure \ref{fig:2wallsp3} ($2$-wall of $Sp(3)$) and Figure \ref{fig:3wallsp5} ($3$-wall of $Sp(5)$), where we also specify the coefficients $\epsilon_i$ of the deformations. 

With this deformation,  we find the expected solutions where the $F$ eigenvalues split into $J$ plus $F-J$. However, we also find additional unexpected solutions, splitting the eigenvalues into more than two different sets. So the global symmetry $Sp(F)$ can be broken to a product of many smaller $Sp$ factors. 

In order to find the moduli space of such solutions, in principle we need to quotient the explicitly broken flavor group (which is a product of many $Sp$ factors) with the sub-group preserved by the eigenvalues trajectories. Doing so, we do not automatically recover the Grassmannians. However let us discuss a possible way of obtaining the Grassmannians.

For instance, in the case of the $1$-walls of $Sp(1)$ with $3$ flavors, we find three solutions of the deformed equations:
\begin{itemize}
\item One solution (left of Figure \ref{fig:1wallsp1}) has explicit global symmetry $Sp(2) \times Sp(1)$, which is preserved by the solution, so its moduli space is trivial (Witten Index $\pm 1$).
 \item Another solution is the charge-conjugated of left of Figure \ref{fig:1wallsp1}, so its moduli space is trivial (Witten Index $\pm 1$).
\item The solution on the right of Figure \ref{fig:1wallsp1} has explicit global symmetry $Sp(2) \times Sp(1)$, broken to $Sp(1)\times Sp(1)\times Sp(1)$ by the eigenvalues trajectories. The moduli space is $\frac{Sp(2)}{Sp(1)\times Sp(1)}=\mathrm{HGr}(1,2)$, (Witten Index $\pm \binom{2}{1}$).\footnote{More generally the moduli space of these solutions include, as factors, \emph{flag manifolds}:
\begin{equation}
\mathcal{F}_{\{k_1,\dots,k_i\}}=\frac{Sp(M)}{Sp(k_1)\times \cdots \times Sp(k_i)}, \quad \sum_{j=1}^i k_j=M\,,
\end{equation}
The WI of such manifolds is given by the formula
\be
WI(\cF_{\{k_1,\dots k_i\}})=\frac{M!}{k_1!\cdots k_i!}.
\ee
}
\end{itemize}
In this case, $Sp(1)$ with $3$ flavors, there is a simple way of organizing the three deformed solutions into the two expected solutions (that is a trivial moduli space with $WI=-1$ and a $\mathrm{HGr}(1,3)$ with $WI=+3$). We combine one trivial solution ($WI=+1$) with a $\mathrm{HGr}(1,2)$ ($WI=+2$) together, to get the $\mathrm{HGr}(1,3)$ expected solution, while the other trivial solution ($WI=-1$) provides the expected trivial solution.

Unfortunately, we do not have a complete analysis of this kind for the parity-invariant wall of $Sp(N)$ for generic $N$. We leave this issue to future work.


\section{Living on the walls}\label{3dTH}

In this section, we discuss the $3d$ effective theories that describe the low energy behavior of the domain walls. 

Such a purely $3d$ description exists  because the vacua of massive $4d$ SQCD develop a mass gap, due to strong interactions and so the $4d$ dynamics below the strong scale $\Lambda$ is trivial. This in turn tells us that, in presence of a domain wall, the degrees of freedom of the theory below the energy scale $\Lambda$  are frozen on the domain wall, which is described by a 3d system decoupled from the 4d bulk.

 The theory living on the domain wall has  3d $\mathcal{N}=1$ supersymmetry, because the domain walls we are considering do not break completely the 4d $\mathcal{N}=1$ supersymmetry, but preserves two supercharges. Moreover, there is a universal part of this 3d theory, described by a 3d free chiral field. The chiral field is composed of a boson, that describes the position of the domain wall along the transverse spatial direction, and its fermionic partner, which is the goldstino of the two supercharges that are broken by the domain wall. We will assume that this part of the 3d theory is always there and in the following we will omit it.

The world-volume theories follow some general requirements. 

The flavor symmetry of the world-volume $3d$ theory matches the symmetry of the (massive) $4d$ theory, which is $Sp(F)$. 

There is a free parameter in the $3d$ theory and tuning such a parameter we end up in different massive phases. This comes about because the 4d model has different IR descriptions depending on the mass parameter of the quarks:  if  $\abs{m_{4d}}\gg\Lambda$, the low energy description of the 4d theory is pure $\mathcal{N}=1$ SYM, instead if $\abs{m_{4d}}\ll\Lambda$ the low energy model is the Wess-Zumino model we discussed in the previous sections.  Therefore, we request that the different phases of 3d world-volume theory describe the different domain walls we found in the two different IR four-dimensional descriptions. 

Another important feature  is that the $3d$ theory for the $k$-wall is IR dual to the theory for the $N+1-k$-wall, up to a parity transformation. Indeed the $k$-wall sector is related to the $N+1-k$ sector because the $k$ vacuum and $N+1-k$ vacuum are related in $4d$ by parity and R-symmetry transformation.

 \subsection{$Sp(N)$ with $F=N+1$}\label{SPN+1}
 
 Let us start from the bulk theory $4d$  $Sp(N)$ with $F=N+1$ flavors, whose global symmetry is $SU(2N+2) \times U(1)$ at zero mass and $Sp(N+1)$ at non-zero mass (which is the situation of interest for us).
 
  We propose that the $3d$ theory living at low energy on the $k$ domain wall is
 \begin{equation}
 \label{eq:SPNN+1 effective theory}
\text{3d} \quad Sp(k)_{\frac{N-k+2}{2}}^{\mathcal{N}=1} \quad \text{ with $N+1$ fundamentals} \quad X.
 \end{equation}
 The fields $X$ are in the fundamental representation of the gauge group and are denoted by the matrix $X_{aI}$, where $a=1,\dots,2k$ is the gauge index and $I=1,\dots,2N+2$. We impose the reality condition $X_{aI}=\Omega^{ab}\Omega^{IJ}X^{\ast}_{bJ}$. In this representation the $Sp(N+1)$ flavor symmetry is manifest. Gauge invariants are constructed in terms of $X_{IJ}^2=X_{aI}X_{bJ}\Omega^{ab}$. $X_{IJ}^2$ is manifestly skew-symmetric. Notice that the $4d$  $Sp(N)$ theory has $2F$ massive fundamentals, while the $3d$ $Sp(k)$ theory has $F$ fundamentals. The $\cN=1$ superpotential is 
\begin{equation}
\mathcal{W}=\frac{1}{4}\Tr( X^2\Omega X^2\Omega)+\frac{\alpha}{4}\Tr(X^2\Omega)^2+m\Tr X^2\Omega\,, \quad
\end{equation}
the SCFT  being at zero mass, $m=0$.

We  assume that a fixed point of the RG flow exists in the region where $\alpha>-\frac 1k$. The overall scale of the superpotential has been fixed for convenience and,  for the sake of studying the model's vacuum structure,  we can set $\alpha=0$. The $3d$ parameter $m$ is the effective IR mass and is related to the $m_{4d}$ parameter. The precise relation between these two parameters is not known, but it is not important because we are interested in the two regimes, large and small four-dimensional mass (compared to the strong scale $\Lambda$), which are related to positive and negative three-dimensional mass. As we show below,  this model has two different phases: a single gapped vacuum for $m>0$ and multiple vacua for $m<0$, hence the proposal meets some of the requirements we demanded for our low energy theory. Moreover, the transition between the two phases is smooth, thanks to the unbroken $\cN=1$ supersymmetry.  

Furthermore, in order to fulfill  the demand that the $k$-domain wall sector is the parity reversal of the $N+1$-wall sector, these models should enjoy the following infrared duality:

\be\label{eq:dualityN+1} \ba{c} Sp(k)_{\frac{N-k+2}{2}}^{\cN=1} \, \text{ with $N+1$ fundamentals } X \\ \cW\sim \abs{X}^4 \ea
           \Llra
\ba{c} Sp(N+1-k)_{-\frac{k+1}{2}}^{\cN=1} \, \text{ with $N+1$ fundamentals } X \\ \cW\sim-\abs{X}^4 \ea \ee

We call the theory on the left Theory A, whereas the model on the right Theory B.

Duality \eqref{eq:dualityN+1} is similar to the family of dualities used in \cite{Bashmakov:2018ghn}, however, the parameter $F$ it is the limit of the range of validity of the duality in  \cite{Bashmakov:2018ghn}. Our domain wall studies suggest that the dualities are valid not only for $0<F<N+1$, but also for $F=N+1$.

$\cN=1$ duality \eqref{eq:dualityN+1} is expected to be an $\cN=1$ deformation of the $\cN=2$ duality for $Sp$ gauge group with non-zero CS, found and tested by Willet and Yaakov \cite{Willett:2011gp}:\footnote{This duality is obtained from the well known $3d$ $\cN=2$ Aharony duality \cite{Aharony:1997gp}
\be \ba{c} Sp(N_c) \, \text{w/ $2N_f$ fundamentals} \\ \cW_{\cN=2}=0 \ea
           \Llra
\ba{c} Sp(N_f-N_c-1) \, \text{w/ $2N_f$ fundamentals} \\ \cW_{\cN=2}=  \mu_{ij}\, tr(p^i p^j)  + \mu \, \M \ea \ee
giving real masses to $ 2 N_f - F$ fundamentals. CS levels $\pm (N_f-F/2)$ are generated (with positive/negative sign on the electric/magnetic side) and $\mu$ becomes massive. Changing variables as $N_c=k, N_f=2+N$, we get \eqref{SPN=2dual}.}
 
\be\label{SPN=2dual} \ba{c} Sp(k)^{\cN=2}_{2+N-\frac{F}{2}>0} \, \text{w/ $F$ fundamentals} \\ \cW_{\cN=2}=0 \ea
           \Llra
\ba{c} Sp(N+1-k)^{\cN=2}_{-N-2+\frac{F}{2}} \, \text{w/ $F$ fundamentals} \\ \cW_{\cN=2}=\mu_{ij}tr(p_i p_j) \ea \ee
which holds for any $F>0$ and $k=1,\ldots,N$. One goes from \eqref{SPN=2dual} to \eqref{eq:dualityN+1} turning on a quartic $\cN=1$ superpotential term, and along the way, on the r.h.s., the gauge singlets $\mu_{ij}$ become massive. See \cite{BS:2021b} for an example of such a deformation discussed with more details, in the case of $U(k) \lra U(N-k)$ CSM dualities. Notice that, for generic $k$ and $F$, the $\cN=2$ theories have $U(F)$ global symmetry, which is enhanced to $Sp(F)$ at the end of the RG flow the lands on the $\cN=1$ SCFTs. This means that the $U(F)$-invariant $Sp(F)$-breaking interactions flow to zero. It is not known that the \eqref{SPN=2dual} duality can be deformed to an $\cN=1$ duality for any value of $k, N, F$. If $F \leq N+1$ the semiclassical analysis of the massive vacua match across the $\cN=1$ duality, so it is very likely that such an $\cN=1$ is correct. In Sec. \ref{SPN+2} we extend this story to $F=N+2$. It would be very interesting to investigate more the regime $F>N+1$, where the models are strongly coupled.

In the following we provide further checks of the duality \eqref{eq:dualityN+1}, studying the two different phases. Varying the mass parameter $m$ from positive to negative values, the vacuum structures of Theory A and of Theory B are the same.  The mapping between the right and left side mass parameter is $m \rightarrow\, - \, m$.  

\subsubsection*{Analysis of the massive vacua of Theory A}
Let us discuss the vacuum structure of the theory A. To do so we have to $2\times 2$-block diagonalize the matrix $X^2 \Omega$ using the gauge and flavor symmetry. The entries $\lambda_i$ of the $2\times 2$- antisymmetric blocks are real because we have imposed a reality condition on $X_{aI}$. Note that the maximal rank of the matrix $X^2\Omega$ is $2k$ and therefore the index $i$ runs from $1$ to $k$.   Once we have diagonalized $X^2 \Omega$, the F-term equations for the ``eigenvalues'' are
\begin{equation}
\lambda_i(\lambda_i^2+m)=0\qquad i\in\{0,\dots, k\} . 
\end{equation}
When $m\neq 0$ these equations have $k+1$ solutions that we will parametrize by $J=0,\dots,k$. Each solutions has only $J$ non-vanishing eigenvalues:
\be
\text{solution} \,\, J: \quad \lambda_1^2,\dots,\lambda_J^2=-m, \quad \lambda_{J+1},\dots,\lambda_{k}=0
\ee 
Depending on the sign of $m$, not all these solutions are acceptable. This gives us a different number of vacua for the two phases of the model.

\textbf{$\bullet \,\, m>0$.} Only the vacuum with $J=0$ is acceptable. In such a vacuum we can integrate out massive quarks with positive mass, leaving
\be
\text{3d} \quad Sp(k)_{N-\frac{k-3}{2}}^{\cN=1}.
\ee
This theory is indeed the TQFT describing  the domain walls of pure $Sp(N)$ SYM  \cite{Bashmakov:2018ghn}. This is exactly what we expected to be the behavior of SQCD at large mass, $m_{4d}\gg \Lambda$. This theory has a single supersymmetric gapped vacuum, it is a Topological Quantum Field Theory and its Witten index is 
\be 
\text{WI}=\smqty(N+1\\k).
\ee 

\textbf{$\bullet \,\, m<0$.} In this case all the $k+1$ vacua are acceptable. Therefore the quarks get a VEV and the matrix $X^2\Omega$ can be put in the a $2\times 2$-block diagonal form, with the first $J$ blocks different from zero. This implies that on the vacua the flavor symmetry is broken to
\be
\label{eq:theoryA}
Sp(N+1)\rightarrow Sp(J)\times Sp(N+1-J),
\ee
leading at low energy to a NLSM with target space 
\be \mathrm{HGr}(J,N+1)= \frac{Sp(N+1)}{Sp(J)\times Sp(N+1-J)}\,.\ee
 Also the VEVs of the quarks break the gauge group as $Sp(k)\rightarrow Sp(k-J)$. Therefore for every $J\neq k$ there is a gauge group left in the infrared. All the fermions charged under the unbroken gauge group shift the CS level with a positive contribution or a negative one depending on the sign of the effective mass. These masses  come either from the potential or from the Higgs mechanism. As a result the CS level of the unbroken gauge group is $Sp(k-J)_{-\frac{k-1-J}2}^{\mathcal{N}=1}$. The NLSM and the CS theory are decoupled in the IR, thus the low energy theory on a vacuum labelled by $J$ is
\be
\label{eq:low energy of SPNN+1}
\text{3d}\quad Sp(k-J)_{-\frac{k-1-J}2}^{\mathcal{N}=1}\times \mathrm{HGr}(J,N+1)
\ee

But not all these vacua are supersymmetric, in fact due to non-perturbative effects a $\cN=1$ CS theory has a supersymmetric vacuum only if the CS level $\mathfrak{h}$ and the rank $h$ of the gauge group satisfy $\frac h2<\mathfrak{h}$ \cite{Witten:1999ds}.  In our case this translates into 
\be 
k-J+1<k-1-J, \quad J<k.
\ee
This relation is never satisfied, so the only acceptable vacuum is the one with $J=k$ on which the gauge group is completely broken.  Only in this case the non-perturbative effects due to the strong dynamics of the gauge group  are not present. One should also point out that this supersymmetric NLSM has a Wess-Zumino term, which is conveniently specified by describing the NLSM as an $\cN=1$ $Sp(J)_{\frac{N+1-J}{2}}$  gauge theory coupled to $N+1$ fundamental scalar multiplets getting VEV. The Witten index of the remaining solution is equal to
\be
\text{WI}=\smqty(N+1\\k).
\ee

\subsubsection*{Analysis of the massive vacua of Theory B}

Carrying out a similar procedure we can study the vacuum structure of  Theory B.

\textbf{$\bullet \,\, m<0$.} We get that there is only one vacuum on which lives a CS theory 
\be
Sp(N+1-k)_{-1-\frac{N+k}{2}}.
\ee
This model is the level-rank dual of the model \eqref{eq:theoryA} an its Witten index is $WI=\smqty(N+1\\N-k+1)$.

\textbf{$\bullet \,\, m<0$.} In this case there are $N+1-k$ vacua. On these vacua the flavor symmetry is broken to $Sp(N+1)\rightarrow Sp(N+1-J)\times Sp(J),$ whereas the gauge symmetry is broken to $Sp(N+1-k)\rightarrow Sp(N+1-k-J)$. As above, integrating out the fermion charged under the unbroken gauge group we shift the CS level. We obtain the low energy theory
\be
Sp(N+1-k-J)_{-\frac{k+1}{2}+\frac{N+1-J}{2}}^{\mathcal{N}=1}\times \mathrm{HGr}(J,N+1)
\ee
 It seems that the number of vacua does not match. But, taking into account the non-perturbative effects, we need to impose $N+2-k-J<N-k-J$. This relation is never satisfied, so the only acceptable vacuum is the one with $J = N+1-k$, on which the gauge group is completely broken.  The only surviving vacuum has WI$=\smqty(N+1\\N+1-k)$ and it matches the vacuum with $m<0$ in theory A because $\mathrm{HGr}(k,N+1)=\mathrm{HGr}(N+1-k,N+1)$.

We stress again that in both models at $m=0$ there is a second order phase transition. The nature of this transition is necessarily second order  and therefore it is described by a $\cN=1$ SCFT. These two SCFTs are the one dual to each other.


\bigskip

A summary of the vacua is displayed in Table \ref{tab:low energy SPN N+1}.
\begin{table}
 \begin{center}
\begin{tabular}{|c|l|c|}
\hline
Wall & Effective theory& Witten Index\\
\hline
$k$& $\mathrm{HGr}(k,N+1)$& $ \smqty(N+1\\k)$\\
\hline
\end{tabular}
\end{center}
\caption{\label{tab:low energy SPN N+1}IR models on the vacuum structure of \eqref{eq:SPNN+1 effective theory} for $m<0$. The various contributions to the Witten index are displayed. }
\end{table}

The vacua we have found in Table \ref{tab:low energy SPN N+1} match precisely the domain wall solutions found in the previous section in Table \ref{table DW theories Sp F=N+1} for the NLSM which describes the low energy regime when $m_{4d}\ll \Lambda$. These models are good candidates for the effective theories on the domain walls of SQCD with gauge group $Sp(N)$ and $N+1$ flavors. 

\subsection{$Sp(N)$ with $F=N+2$}\label{SPN+2}

Let us now add one flavor, considering the $4d$ theory $Sp(N)$ with $F=N+2$.  We propose that the  low energy theory describing the low energy behavior of the  $k$-wall is
\begin{equation}
\label{eq:eff theory N+2}
Sp(k)_{\frac{N-k+1}{2}}^{\mathcal{N}=1} \quad \text{with } N+2 \quad \text{fundamentals}\,\,\, X.
\end{equation}
In this case the amount of evidence that we can provide is weaker than for $Sp(N)$ with $F \leq N+1$ flavors, or for $SU(N)$ with $F<N$ flavors. We still have the $\cN=2$ duality \eqref{SPN=2dual} 
\be\label{SPN=2dualN+2} \ba{c} Sp(k)^{\cN=2}_{1+\frac{N}{2}>0} \, \text{w/ $N+2$ fundamentals} \\ \cW_{\cN=2}=0 \ea
           \Llra
\ba{c} Sp(N+1-k)^{\cN=2}_{-1-\frac{N}{2}} \, \text{w/ $N+2$ fundamentals} \\ \cW_{\cN=2}=\mu_{ij}tr(p_i p_j) \ea \ee
from which we expect to get an $\cN=1$ duality incarnating the equivalence of a $k$ wall with the time-reversed $N+1-k$ wall. This duality is
\be\label{SPN=1dualN+2} \ba{c} Sp(k)^{\cN=1}_{\frac{N-k+1}{2}>0} \,\\ \text{with $N+2$ fundamentals}  \ea
           \Llra
\ba{c} Sp(N+1-k)^{\cN=2}_{-\frac{k}{2}} \, \\\text{with $N+2$ fundamentals,}  \ea \ee
we will call the theory on the left Theory A, whereas we will call Theory B the model on the right.

\subsubsection*{Analysis of the massive vacua of Theory A}

Let us now study the vacuum structure of \eqref{eq:eff theory N+2}.
The superpotential we are considering is the same we have considered in the case with $F=N+1$, namely
\begin{equation}
\label{eq:supN+2k}
\mathcal{W}=\frac{1}{4}\Tr( X^2\Omega X^2\Omega)+\frac{\alpha}{4}\Tr(X^2\Omega)^2+m\Tr (X^2\Omega).
\end{equation}
Here again we assume that  an RG flow fixed point exists in the region with $\alpha>-\frac1k$, and henceforth we set $\alpha=0$. The vacua analysis is similar to the one we have done for the $F=N+1$ case and so we recall schematically what we found there.
After the block diagonalization, the F-term equations read
\be 
\lambda_i(\lambda_i^2+m)=0\qquad i\in\{0,\dots, k\} 
\ee

These equations have $k+1$ solutions, which will be parametrized by $J=0,\dots,k$, but depending on the sign of $m$, not all are acceptable.

\textbf{$\bullet$ $m>0$.} There is only one solution $J=0$ and the low energy theory is the TQFT 
\be
\label{eq:AVdualSp}
\text{3d} \quad Sp(k)_{N-\frac{k-3}{2}}^{\cN=1},
\ee
once we have integrated out the positive mass fermions.

\textbf{$\bullet$ $m<0$.} In this case all $J$ solutions are acceptable. Therefore the quarks $X^2\Omega$ take VEVs and break both the flavor symmetry, $Sp(N+2)\rightarrow Sp(N+2-J)\times Sp(J)$, and the gauge symmetry $Sp(k)\rightarrow Sp(k-J)$. The low energy models living on each of the $J$ vacua are
\be
Sp(k-J)_{\frac{k-J+1}{2}}^{\cN=1}\times \mathrm{HGr}(J, N+2).
\ee
Contrary to the $F=N+1$ case, here the strong dynamic does not break supersymmetry because the CS factor of the effective theory is such that there is a unique supersymmetric vacuum, with trivial TQFT. Therefore all the vacua are supersymmetric vacua. Computing the Witten index is now a subtle task because as pointed out by \cite{Bashmakov:2018wts} the sign between the $J$ vacua depend on the number of charged fermions with negative mass. After a careful analysis of the masses  of the charged fermions under the gauge group, generated either by the superpotential or via Higgs mechanism, we get that the Witten index is given by
\be
\text{WI}=\sum_{J=0}^k(-1)^{J+k}\smqty(N+2\\J).
\ee

\bigskip

\subsubsection*{Analysis of the massive vacua of Theory B}

In this model we consider the superpotential
\begin{equation}
\label{eq:supN+2kB}
\mathcal{W}=-\frac{1}{4}\Tr( X^2\Omega X^2\Omega)-\frac{\alpha}{4}\Tr(X^2\Omega)^2+m\Tr (X^2\Omega),
\end{equation}
and again we consider $\alpha>-\frac 1k$, hence setting $\alpha=0$ for simplicity. The study of the vacua goes on as we have done in the previous subsection, and we are going only to list the different phases. 

\textbf{$\bullet$ $m<0$.} There is only one vacuum and the low energy theory is the TQFT 
\be
\label{eq:AVdualSp}
\text{3d} \quad Sp(N+1-k)_{-\frac{k+N+2}{2}}^{\cN=1},
\ee
once we have integrated out the negative mass fermions.

\textbf{$\bullet$ $m>0$.} In this case there are $H=0,\dots N+1-k$ vacua.  The quarks $X^2\Omega$ take VEVs and break both the flavor symmetry, $Sp(N+2)\rightarrow Sp(N+2-H)\times Sp(H)$, and the gauge symmetry $Sp(N+1-k)\rightarrow Sp(N+1-k-H)$. The low energy models living on each of the $H$ vacua are
\be
Sp(N+1-k-H)_{\frac{N+2-k-H}{2}}^{\cN=1}\times \mathrm{HGr}(H, N+2).
\ee

This analysis of the massive vacua in some cases provides more vacua than the $4d$ analysis. The vacua expected are always there, but in some cases there are more vacua. Indeed, only when the rank of the gauge group $k\leq\frac{N+1}{2}$ the vacua match the $4d$ domain wall solutions for the Theory A, while when the gauge group $k > \frac{N+1}{2}$ the $3d$ theory presents more vacua at large negative masses. The behaviour of Theory B is the opposite: when $k\leq\frac{N+1}{2}$ there are too many vacua, while when $k > \frac{N+1}{2}$ the number of vacua matches the $4d$ analysis. Moreover, the vacua do not match perfectly across the $\cN=1$ duality.

We interpret this mismatch as follows: our interpretation of the semiclassical analysis of the $3d$ vacua is  naive when the $3d$ theories under consideration are strongly coupled. They might present \emph{quantum phases}  similar to what happens for non-supersymmetric theories: for fixed rank and number of flavors, if the  Chern-Simons level is too small there are \emph{quantum phases} and a naive analysis of the vacua is incorrect.

\bigskip

 To sum up the vacua for $m<0$ are reported in Table \ref{tab:low energy SPNN+2}.
 
 \begin{table}[h!]
 \begin{center}
\begin{tabular}{|c|l|c|}
\hline
Wall& Effective theory& Witten Index\\
\hline
$k$& $\mathrm{HGr}(J,N+2),\quad J\in \{0,\dots,k\}$&  $ \smqty(N+1\\k)=\sum_{j=0}^{k}(-1)^{j+k}\smqty(N+2\\j)$\\
\hline
\end{tabular}
\end{center}
\caption{\label{tab:low energy SPNN+2} IR models on the vacuum structure of \eqref{eq:eff theory N+2} for $m<0$ and $k\le\frac {N+1}2$. The various contributions to the Witten index are displayed. }
\end{table}
So long as $k \leq \frac{N+1}2$, this Table \ref{tab:low energy SPNN+2} exactly reproduces the Table \ref{table DW theories Sp F=N+2} of solutions we have found for the WZ model which describes SQCD when $m_{4d}\ll \Lambda$.

In the case of the parity-reversed wall $k=(N+1)/2$, we expect the infrared duality
\be Sp(k)_{\frac{k}{2}}^{\mathcal{N}=1} \quad \text{with } 2k+1  \quad \lra \quad  Sp(k)_{-\frac{k}{2}}^{\mathcal{N}=1} \quad \text{with } 2k+1  \,,\ee
which tells us that the infrared SCFT is $3d$ parity invariant. The parameters in these cases are such that  strong-coupling effects do not play an important role (this is because the absolute value of the CS level of the $Sp(N_c)$ theory is equal to, not smaller than, $\frac{N_c+1}{2}$). 


\section*{Acknowledgements}
We are very grateful to Francesco Benini for initial collaboration, useful suggestions and careful reading of the manuscript.

\bibliographystyle{ieeetr}

\bibliography{Non_SUSY_dualities}

\end{document}